\documentclass[12pt,preprint]{aastex62}
\usepackage{xspace}

\newcommand{\cchp}{\,CCHP\xspace}

\newcommand{\ho}{$H_{0}$\xspace}                  
\newcommand{\hounits}{\,km\,s$^{-1}$\,Mpc$^{-1}$\xspace}   
\newcommand{\hst}{\emph{HST}\xspace}

\newcommand{\jwst}{\emph{JWST}\xspace}

\newcommand{\sne}{SNe~Ia\xspace}

\shorttitle{JAGB Distance Scale}
\shortauthors{Freedman \& Madore}

\usepackage{cancel}

\begin{document}


\title{\bf Astrophysical Distance Scale \\ II. Application of the JAGB Method:
\\ A Nearby Galaxy Sample}



\author{\bf Wendy L. Freedman}
\affil{Dept. of Astronomy \& Astrophysics \\ University of Chicago \\ 5640 S. Ellis Ave.,
\\ Chicago, IL, ~~60637} 
\email{wfreedman@uchicago.edu}

\author{\bf Barry F. Madore} 
\affil{Dept. of Astronomy \& Astrophysics \\ University of Chicago \\ 5640 S. Ellis Ave.,
\\ Chicago, IL, ~~60637} 
\affil{The Observatories \\ Carnegie
Institution for Science \\ 813 Santa Barbara St., Pasadena, CA ~~91101}


\begin{abstract}\medskip
\medskip
We apply the near-infrared $J$-region Asymptotic Giant Branch (JAGB) method, recently introduced by Madore \& Freedman (2020), to measure the distances to 14 nearby galaxies out to 4~Mpc. We use the geometric detached eclipsing binary (DEB) distances to the  LMC and SMC as independent zero-point calibrators. We find excellent agreement with previously published distances based on the Tip of the Red Giant Branch (TRGB): the  JAGB distance determinations (including the LMC and SMC) agree in the mean to within $\Delta(JAGB-TRGB) = $+0.025~$\pm$ 0.013~mag, just over 1\%, where the TRGB $I$-band zero point is $M_I = $~-4.05~mag. 
With further development and testing, the JAGB method has the potential to provide an independent calibration of Type Ia supernovae (\sne), especially with $JWST$.  The JAGB stars (with $M_J = -6.20$~mag) can be detected farther than the fainter TRGB stars, allowing greater numbers of calibrating galaxies for the determination of \ho. Along with the TRGB and Cepheids, JAGB stars are amenable to theoretical understanding and further refined empirical calibration. A preliminary test shows little dependence, if any, of the JAGB magnitude with metallicity of the parent galaxy. These early results suggest that the JAGB method has considerable promise for providing high-precision distances to galaxies in the local universe 
that are independent of distances derived from the Leavitt Law and/or the TRGB method; and it has numerous and demonstrable advantages over the possible use of Mira variables.
\end{abstract}


\medskip

\section{Introduction}
The recent discrepancy between the values of the Hubble constant measured locally (based on \sne calibrated by Cepheids), and that inferred at high redshift from measurements of the CMB, has been well-documented (e.g., Freedman et al. 2012; Riess et al. 2016; Planck collaboration  2016, 2018; Zhang et al. 2017; Freedman 2017; Feeney et al. 2018; Riess et al. 2019; Freedman et al. 2019). Wong et al. (2019) have used time delay measurements for strong gravitational lenses and conclude that the discrepancy is at a 5.3-$\sigma$ level. (However, see Kochanek 2020 who concludes that it is unlikely that any recent estimates from gravitational lens time delays are accurate to better than 10\%). Lower values of \ho, consistent with the Planck value, come from studies of baryon acoustic oscillations combined with \sne (where the scale is set by the sound speed at recombination) (e.g., Aubourg et al. 2015; Macaulay et al. 2019). Additional, independent calibrations of the zero point to the extragalactic distance scale are therefore essential to test whether unknown systematics could be entering into the local Cepheid calibration and contributing to the discrepancy.

The Tip of the Red Giant Branch (TRGB) offers an alternative and independent route to measuring the distances to nearby galaxies: this method has high precision and high accuracy, comparable to the Cepheids (see Lee et al. 1993; Jang \& Lee 2017; Hatt et al. 2017; Freedman et al. 2019). The Carnegie-Chicago Hubble Program (\cchp) has recently used the TRGB method to provide a new calibration of \ho (Freedman et al. 2019). The value of \ho based on this TRGB calibration lies midway between those obtained at low and high redshifts, yielding \ho = 69.8 $\pm$ 0.8 ($\pm$1.1\% stat) $\pm$ 1.7 ($\pm$2.4\% sys) \hounits (see also Freedman et al. 2020).

In this paper, we explore yet another, very promising technique for measuring high-precision, accurate distances to nearby galaxies, a method that has not, to date, been developed in the context of measuring absolute distances to galaxies.  We analyze previously published data for a class of asymptotic giant branch (AGB) carbon stars that show exciting promise as an additional, independent means for calibrating the local value of \ho. We note that oxygen-rich Mira variables offer another, but significantly more challenging route to measuring the distances to nearby galaxies (e.g., Feast 2004; Whitelock et al. 2008; Huang et al. 2018.) Later, in \S\ref{sec:Miras}, we will discuss in more detail the pros and cons of Mira variables as extragalactic distance indicators.

\begin{figure}[hbt!] 
 \centering
\includegraphics[width=12.5cm, angle=-00]{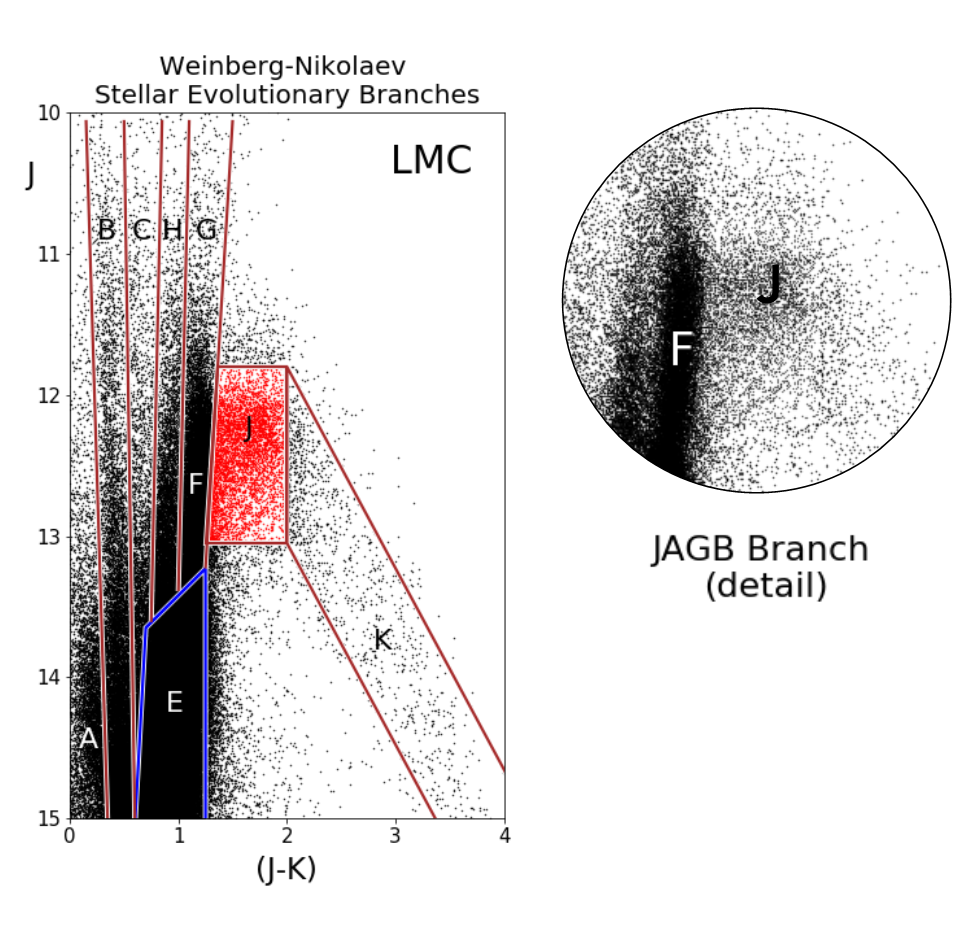}
\caption{\small  The Weinberg-Nikolaev stellar-population branches labeled in a near-infrared $J$ vs $(J-K)$ color-magnitude diagram; 
based, in this case, on the data of Macri et al. (2015). Of interest here is Region $J$, centrally located, shown in red and marked with a letter `J'. The oxygen-rich $AGB$ population is marked  with an `F'; while the Red Giant Branch is outlined in blue below it, and designated by an `E'. A blow-up of the $J$ Region is shown in the circular area to the right in this figure. As discussed in MF20, in the near-infrared $J$-band, the JAGB stars have a constant magnitude. 
} 
\label{fig:LMC}
\end{figure}

This class of AGB stars was labeled by Nikolaev \& Weinberg (2000) and Weinberg \& Nikolaev (2001;  hereafter, WN01) as one of a number of  regions of  stellar populations in the near-infrared (2MASS) color-magnitude diagrams of the Large Magellanic Cloud (LMC). In particular, they drew attention to a large and distinctive group of very red and very bright stars with a low dispersion of their absolute magnitudes in the observed near-infrared bands. They designated these stars as being members of 
``Region J", one of eleven of their empirically-defined sequences of stars (see Figure \ref{fig:LMC}).  These stars were found to brighten in magnitude (in the $K$-band) linearly with increasing color (WN01 used the $K$-band and $(J-K)$ color). For the purposes of using these stars for measuring distances, we will hereafter refer to these stars J-region stars as `JAGB' stars.\footnote{ We note that a sparsely-populated tail redward of Region $J-$ (designated Region $K$ stars in WN01 and in Figure \ref{fig:LMC} here) systematically drops quite rapidly in luminosity with color. These stars are so rare that for our purposes (extragalactic distance determinations) they are of little use, especially in comparison to the JAGB stars being studied here, and they are easily eliminated by a simple red color cut.} WN01 then went on to use these JAGB stars to successfully map out the {\it differential} back-to-front geometry of the LMC.

In our first paper on this topic (Madore \& Freedman 2020; hereafter MF20) we used the JAGB method to measure the distance to the Sculptor group galaxy NGC~253.  
For this galaxy, we utilized the JAGB stars as {\it absolute} rather than {\it relative} distance indicators, calibrating them through both the LMC and the SMC. NGC~253 is at a distance of 3.4~Mpc, about seven times farther away than the LMC, and well outside of the Local Group. In this paper, we apply the method to 14 additional nearby galaxies, and undertake a comparison of the distances obtained with the JAGB technique to those measured and previously published using the TRGB method.

\section{JAGB Stars in Context}
\subsection{Historical}
While not as commonly considered as are Cepheids or the $TRGB$, the use of carbon stars for measuring extragalactic distances is, in fact, not new. Before turning to our results applying the JAGB method to the distance scale, we provide here a brief summary of previous  applications of carbon stars, in specific, to the local distance scale. 

Almost 40 years ago, Richer (1981) undertook a broad survey of carbon stars in the LMC using optical $VRI$ data. He found that the carbon-star bolometric luminosity function was  bright (-5.0 mag), strongly peaked, symmetric, and moderately well defined (with $\sigma \sim \pm$~0.45 mag; his Figure 5). Later, Richer, Crabtree \& Pritchet (1984) discovered seven carbon stars in the Local Group dwarf elliptical galaxy, NGC~205, using a combination of narrow and broad-band imaging techniques. Their technique made use of the fact that carbon stars have deep CN bands at $\sim$8100\AA ~and no TiO, whereas M stars have TiO absorption at $\sim$7800\AA ~and no CN. Comparing the I-band photometry for the seven carbon stars to the I-band luminosity function for the aforementioned LMC sample, they took the first step in establishing a carbon-star-based (extragalactic) distance scale. Richer, Pritchet \& Crabtree (1985), Pritchet et al. (1987) and Richer \& Crabtree (1985) extended these studies to NGC 300, NGC 55 \& M31, respectively. 

Soon thereafter, Cook, Aaronson \& Norris (1986) undertook a  (narrow-band) photometrically-selected survey for carbon stars in five Local Group galaxies (M31, M33, NGC~6822, IC~1613 and WLM). They too noted the similarity of the mean magnitudes of the $I$-band luminosity functions, despite the small samples available to them. At that time, independently-determined distances to most of these galaxies were still not known to better than 10\%; nevertheless, they suggested that these stars could potentially become useful distance indicators.

The most recent review of the status of carbon stars as distance indicators was written 15 years ago by Battinelli \& Demers (2005), entitled ``The Standard Candle Aspect of Carbon Stars”. They summarized the growing evidence that carbon stars had a small dispersion in their $I$-band magnitude,  and that the mean luminosity of these stars might also be relatively constant from galaxy to galaxy. Battinelli \& Demers ended their review by noting, even at that early juncture, that \jwst could provide unique opportunities for extending the distance scale using carbon stars.

As we shall see in \S4 and \S6 (below), the near-infrared luminosity function for JAGB stars exhibits an even smaller dispersion in the $J$ band than the $I$ band (used in the studies described above). In addition, the $J$ band has a nearly constant mean absolute magnitude (with color). Following on the work of WN01 and their identification of a ``Region J ", we show that these stars are easily identified and distinguished on the basis of their $(J-K)$ or $(J-H)$ colors. We have surveyed the published literature for galaxies with near-infrared photometry that are located within the Local Group and out to distances of $\sim$4~Mpc. These galaxies also have accurate and independently-determined distances, measured using the TRGB. As such they form a homogeneous set of distances that can be compared to the JAGB distances. We emphasize that these previously published data were not obtained (nor optimized) for the purpose of measuring distances to these galaxies. However, the excellent agreement with previously published distances to these galaxies (as presented in  \S\ref{sec:compare}) suggests that, with optimal observing strategies, and the extension to the Hubble Space Telescope (\hst), the James Webb Space Telescope (\jwst) and {\it WFIRST}, this method has enormous potential for improving and strengthening the local extragalactic distance scale and then providing a calibration of the expansion rate of the universe, \ho.

\subsection{Modeling of AGB Stars}
\label{sec:theory}

 In the course of their evolution, AGB stars undergo oscillations or thermal pulsations. As a result, carbon can be brought to the surface, particularly  during the third and later (generally deeper) dredge-up phases (Iben \& Renzini 1983;  Herwig 2013; Habing \& Olofson 2004 and references therein), leading to the formation of carbon stars. Not all AGB stars become carbon stars: in a carbon star, the atmosphere contains more carbon than oxygen, in a ratio of C/O $>$1.  Recent modeling of AGB star evolution has been carried out by many authors, including Marigo et al. (2008, 2017), Karakas (2014); Rosenfield et al. (2014); Salaris et al. (2014); and Pastorelli et al. (2019).

 An illustrative description that ties directly to the various regions characterized by Nikolaev \& Weinberg (2000) and WN01 (Figure \ref{fig:LMC} above), is given by  Marigo et al. (2008), and summarized below:

Younger (more massive) AGB stars with ages of around $1.6 \times 10^8$ years correspond to  Region G  in Figure \ref{fig:LMC} above.  For these luminous AGB stars, the temperature at the  bottom of the convective envelope is hot enough that convective-envelope burning (also known as hot bottom burning, HBB) can occur. The HBB converts the carbon into nitrogen, before it can reach the surface, and thereby prevents these (and all more massive) AGB stars from having carbon-enhanced atmospheres (Boothroyd et al. 1993). 

For older (less massive) AGB stars (with ages of about $1.3 \times 10^9$ years), the carbon-rich material at the bottom of the convective zone can be mixed into the outer envelope from a region that is lower in temperature wherein HBB does not occur. These stars are found in ``Region J " (as seen in Figure \ref{fig:LMC}), that is populated by carbon stars, the subject of this paper, the JAGB stars.

For even older and less massive AGB stars (ages of $10 \times 10^9$ years, or more), the situation is different yet again; for these stars, there is no third (deep) dredge-up phase. These lower-luminosity stars correspond to Region F in Figure \ref{fig:LMC} above. 

Thus, a combination of both 1) the addition of carbon to the envelope during third dredge up and 2) conditions such that HBB does not occur, gives rise to a population of high-luminosity JAGB (carbon) stars having a small dispersion in luminosity, and the clear potential for use as standard candles in the extragalactic distance scale.

Before closing this section, we note that, as for the case of Cepheids or the TRGB and their application to the extragalactic distance scale, we also take a purely empirical approach to the JAGB distance method.  For example, models cannot yet successfully calibrate the period-luminosity relation for distance measurements to  Cepheids, nor do they calibrate the He flash for the TRGB. As we shall show below,  the JAGB LF appears to exhibit a stable peak (in the J-band) over a well-defined defined range in color. This observed characteristic may be a helpful constraint for AGB models.

\section{The LMC Calibration}
\label{sec:lmc_calib}

As discussed in MF20, the near-infrared $JHK$ data used for the LMC-based zero-point calibration of the JAGB method are those of Macri et al. (2015). (We note that the Macri et al. survey has 3,341 JAGB stars. For comparison, 2MASS sampled the entire galaxy and has in excess of 7,000 JAGB stars.) The reasons for adopting the Macri et al. observations over the $2MASS$ dataset (used by WN01) are given in MF20, but we re-iterate here that the Macri et al. data are of higher precision and of better spatial resolution than the $2MASS$ data. In addition, these data are based on 16 epochs of observing, and are temporally averaged. Finally, the Macri sample is located in the bar of the LMC, where the majority of the detached eclipsing binary (DEB)  stars that are used for a geometric determination of the distance to the LMC, were observed by Pietrzynski et al. (2019). We adopt for the LMC bar, and by association, for our JAGB stars, a true distance modulus of $\mu_o = $ 18.477 $\pm$ 0.004 (stat) $\pm$ 0.026 (sys), as measured by Pietrzynski et al. 

On the left side of Figure 2 we show the $J$ versus $(J-H)$ color-magnitude diagram (CMD) for the LMC. On  the right side we show the 
JAGB luminosity function, extracted from the  color cut made in the CMD, as denoted by the two vertical
lines at $(J-H)$ = 1.10 and 1.35~mag. The mean magnitude ($J = $~12.31~mag) and dispersion ($\pm$~0.21~mag)
for the JAGB population was calculated in the $\Delta J =~ \pm$0.4~mag box,
shown in the CMD. (Changing the box width from $\pm$0.3~mag to $\pm$0.5~mag resulted in a change of $\pm$0.017 mag in the mean J-band magnitude of the JAGB zero point.) A split-dispersion Gaussian, representing the slightly skewed JAGB luminosity function, was then centered on the calculated mean, and scaled to fit the wings, as shown in the right panel. The 3,341 JAGB stars 
in the sample yield an error on the mean of $\pm$ 0.21/$\sqrt{3341} = $~0.004~mag. For a true distance modulus of 18.477~mag, as adopted above, we then have an absolute calibration of the JAGB method, with $M_J = -6.22 \pm 0.004$~(stat)~$\pm$ 0.026~mag (sys). 
The SMC, which also has a DEB distance gives $M_J = $ -6.18 $\pm$ 0.006~(stat) $\pm$ 0.048~mag (sys) (see MF20 for details). Averaging the two results, our provisionally-adopted JAGB zero point, used throughout this paper, 
is $M_J (JAGB) = $ -6.20 $\pm$ 0.037 (sys)~mag.

\vfil\eject

\subsection{2MASS, DENIS and VISTA Systems} 
In future, it is our intention to re-observe and update these individual galaxy distances so that they are
uniformly tied to the 2MASS photometric system.
However, for this paper it has been necessary to consider data obtained in other 
surveys in order to take this first look at the JAGB as a distance indicator.
What concerns us most, at this juncture, is the cross-survey stability of 
the published J-band magnitudes with respect to 2MASS. Before making those
comparisons for three of the other most popular near-IR survey instruments
(VISTA, UKIRT/WFCAM and DENIS) we note that the observed scatter in Figure 5
must, by necessity, already include any scatter introduced by the photometric
zero-point inhomogeneity that may, or may not, exist between the data sets used
here. In other words, once put onto a single photometric system, 
the scatter in the JAGB method is likely to become even lower than it is now.

VISTA: Gonzales-Fernandez et al. (2015) describe ``the routine calibration of data taken with the VISTA camera" where they state that ``the broad-band $JHK_s$ data are directly calibrated from 2MASS point sources visible in every VISTA image." They go on to quantify this statement with the equation, $ J_{VISTA} = J_{2MASS}  - 0.013 (\pm0.006) (J-K_s)$.

UKIRT/WFCAM:  In the same paper just cited above, the authors give further equations that yield $J_{WFCAM} - J_{2MASS} = +0.002 (\pm0.020)$~mag.
Additionally, Carpenter (2003) gives $J_{2MASS} = J_{UKIRT} +0.074(\pm0.015)(J-K)_{UKIRT} -0.012 (\pm0.006)$~mag.

DENIS: Cabrera-Lavers \& Garzon (2004) conclude ``that there are no significant differences between 2MASS and DENIS photometry'', noting, in the caption of their Figure 4, a shift of $J_{MASS} - J_{DENIS} = +0.003 (\pm 0.003$~mag (from their Table 2).

 \begin{figure}[htb!]
\centering
\includegraphics[width=10.5cm,angle=0]{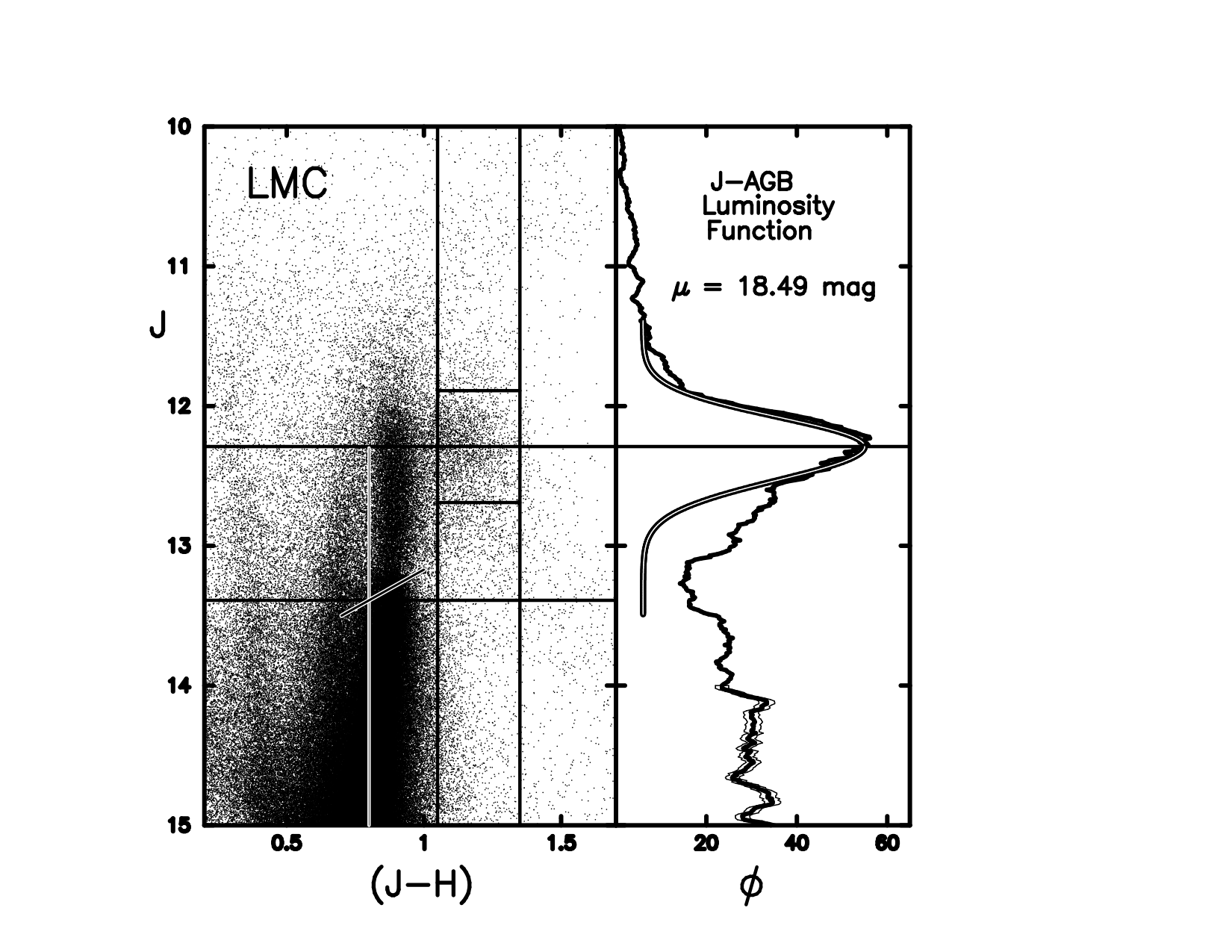} 
\caption{\small  $J$ versus $J-H$ color-magnitude diagram for JAGB Stars in the LMC. The left portion of the figure shows the near-IR color-magnitude diagram for stars 
in the LMC centered on the dominant population of $AGB$ stars (Region \emph{F} in Figure 1) starting at about $J = $ 11.7~mag and continuing down to the tip of the red giant branch at $J = $ 13.3~mag. Region $J$ AGB stars are highlighted here by the solid vertical lines at 1.10 and 1.35~mag in $(J-H)$ color, and by two short horizontal lines at $J = $ 11.9 and 12.7~mag.  The right portion of the panel shows the $J-$band luminosity function marginalized between the two aforementioned color cuts.   The two longer horizontal lines denote the mean JAGB magnitude and the position of the TRGB below it, which is shown in this plot by the slanting white line at J = 13.4 mag. Note the one-magnitude advantage that the JAGB stars have over the TRGB.
} 
\label{fig: LMC2}
\end{figure}

\section{Application of the JAGB Method}   
   
In addition to our first target, NGC~253, beyond the calibrating LMC and SMC (all discussed in MF20), we now apply this method to an extended sample of 13 additional galaxies, for which ground-based, near-infrared photometry has been published and made publicly available. We reiterate that these data were taken for other purposes, and were not optimized for the determination of distances via the JAGB method; and in some cases (e.g., Phoenix), there are very few JAGB stars visible. However, overall this sample provides an excellent starting point to test the viability and precision of the method when compared to other published distances to these same galaxies, using independent methods.

The galaxies in our current sample cover a wide range of intrinsic and extrinsic properties. Given these differences, each of these systems tests the JAGB distance determination method in different ways. For instance, the JAGB population sizes differ considerably (ranging from a low signal-to-noise detection in WLM, to high-precision individual observations of over 3,300 
JAGB stars in  the LMC, and over 1,600  in the SMC, our zero-point calibrator galaxies. (See MF20 for a discussion of the SMC sample.) Their distances (which impact both the degree of crowding encountered and the final signal-to-noise at the level of the JAGB stars) differ by nearly two orders of magnitude. Finally, the total line-of-sight extinction encountered by these galaxies range from extremely small values $(E(B-V) \sim $~0.00 - 0.02 mag) for the SMC, Leo~I and IC~1613, to reddenings up to fifteen times higher (e.g., $E(B-V) \sim$ ~0.35 mag), as in the case of NGC~6822.

In the process of collecting JAGB data on this sample of galaxies, we also compiled measurements of TRGB distances to the same galaxies. However, in order to compile a self-consistent (and non-overlapping) data set, not all of the published TRGB distances were included in this compilation. Our acceptance criteria were simple: Only ground-based I-band (Kron-Cousins) data or its space-based equivalent (e.g., F814W data from \hst) were included; Sloan {\it i-}band data, for example, were not considered.  Each of the original publications were inspected and the measured tip magnitude (explicitly uncorrected for any line-of-sight extinction) was tabulated. Care was taken to avoid double counting measurements because, for instance, in many cases the same data were published, updated, added to compilations and/or republished by the same groups or individuals within those groups. In those cases only the most recently updated/published value was taken. If independent groups published a tip detection using the same data but employing different sample selection and tip measurement techniques those values were also included in our sample. All of the apparent magnitudes of the TRGB for a given galaxy were then uniformly corrected for a single line-of-sight foreground Galactic extinction, tabulated in NED as derived from Schlafly \& Finkbeiner (2011). The extinction-corrected magnitudes were then individually converted into true distance moduli using our adopted $I$-band TRGB intrinsic tip absolute magnitude of $M_I = $ -4.05~mag (Freedman et al. 2020). Each of the references and the self-consistently adopted foreground extinctions and true distance moduli are presented in the discussion sections for each individual galaxy below, averaged and then summarized in Table 1.

\begin{figure}[htb!] 
\includegraphics[width=5.5cm, angle=0.]{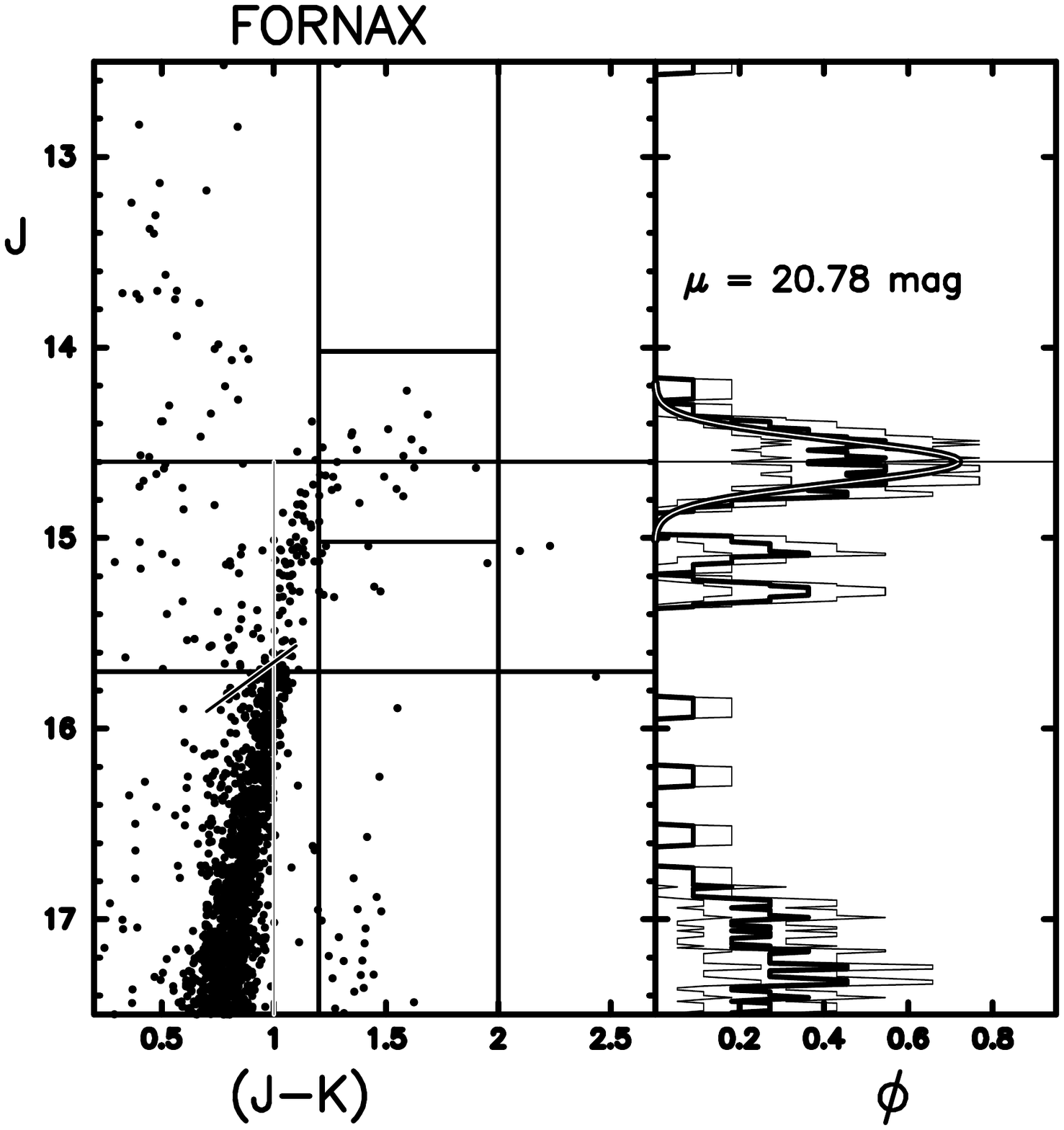} 
\includegraphics[width=5.5cm, angle=0]{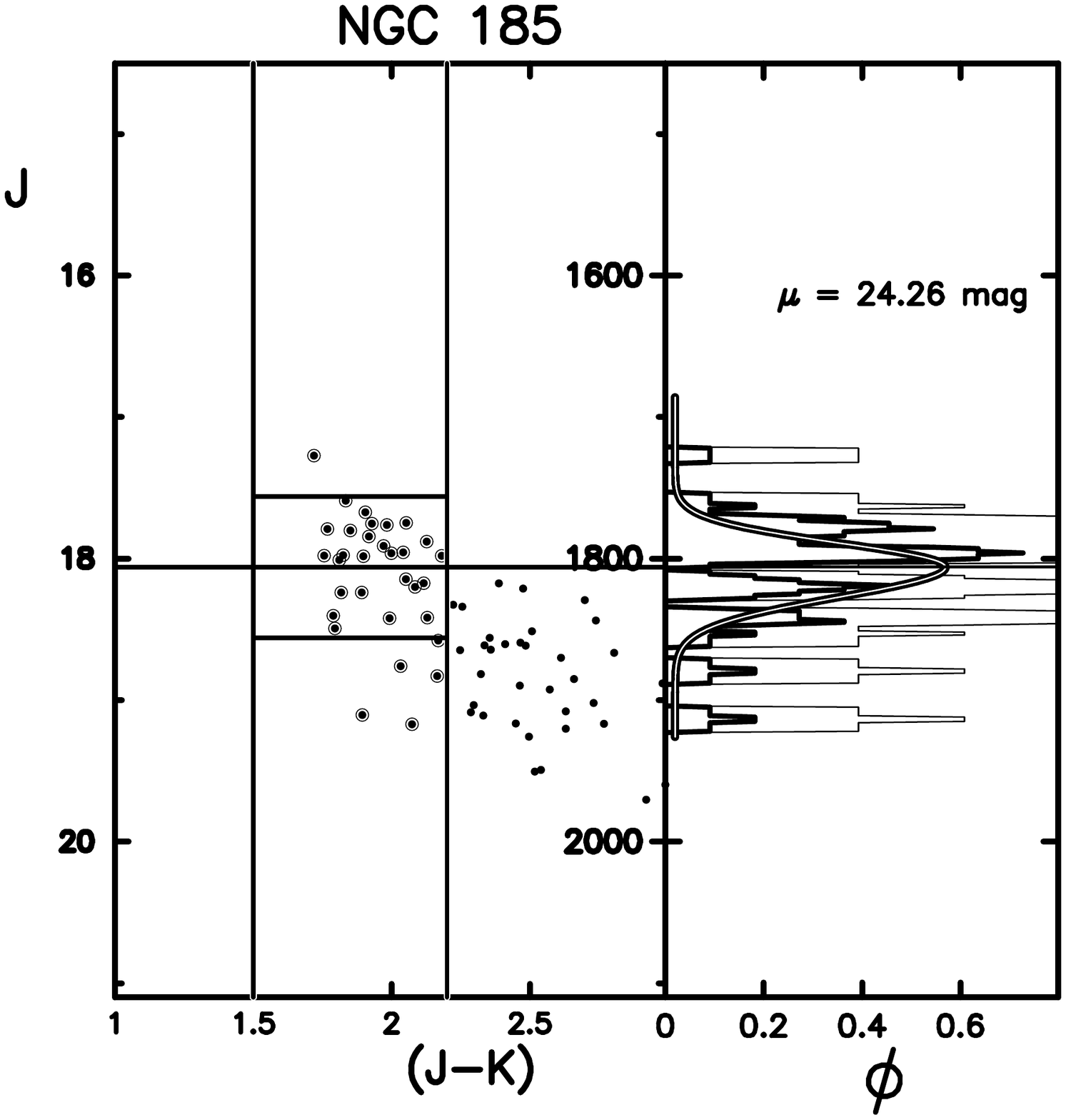} 
\includegraphics[width=5.5cm, angle=0]{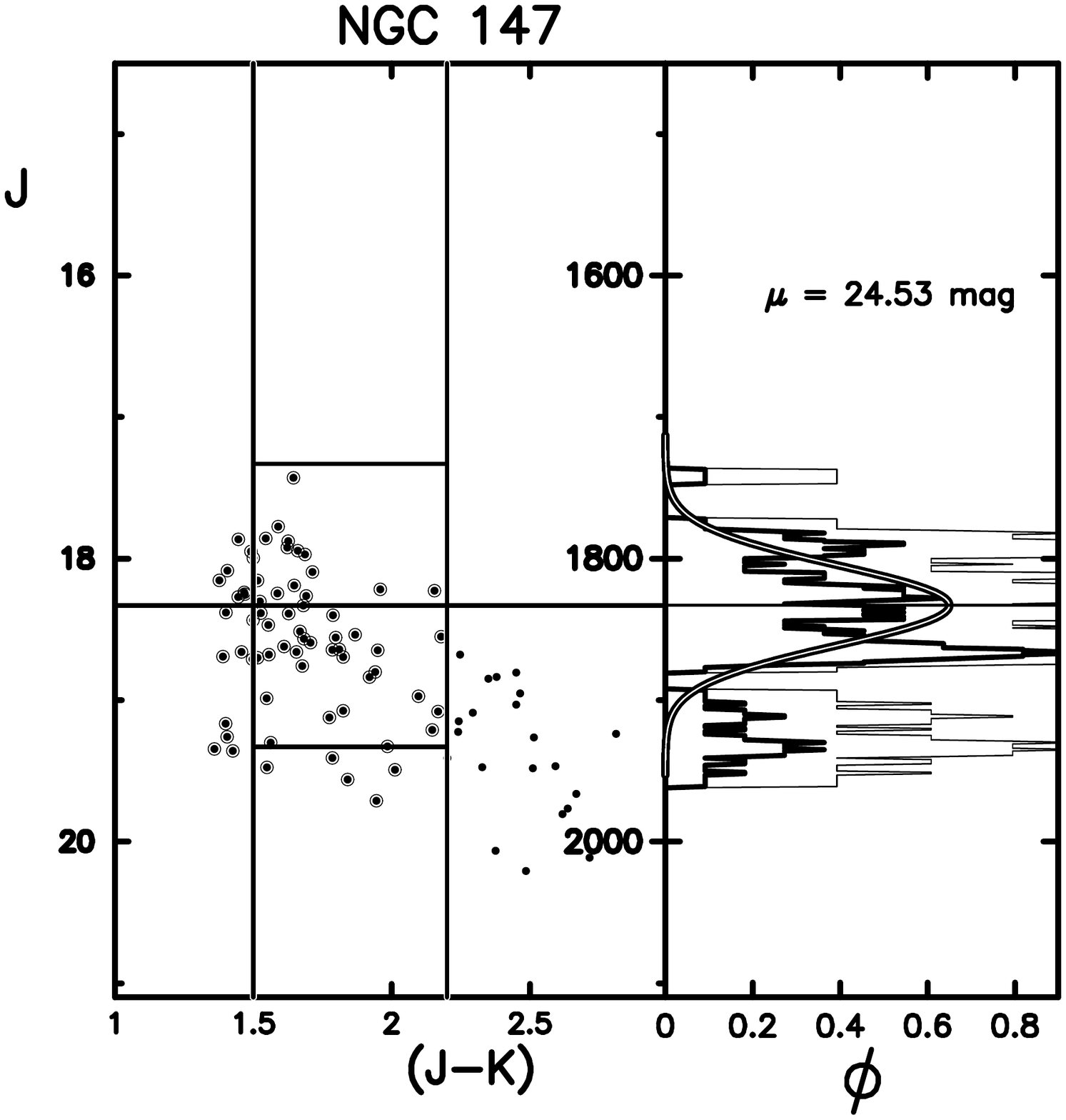} 

\includegraphics[width=5.5cm, angle=-0]{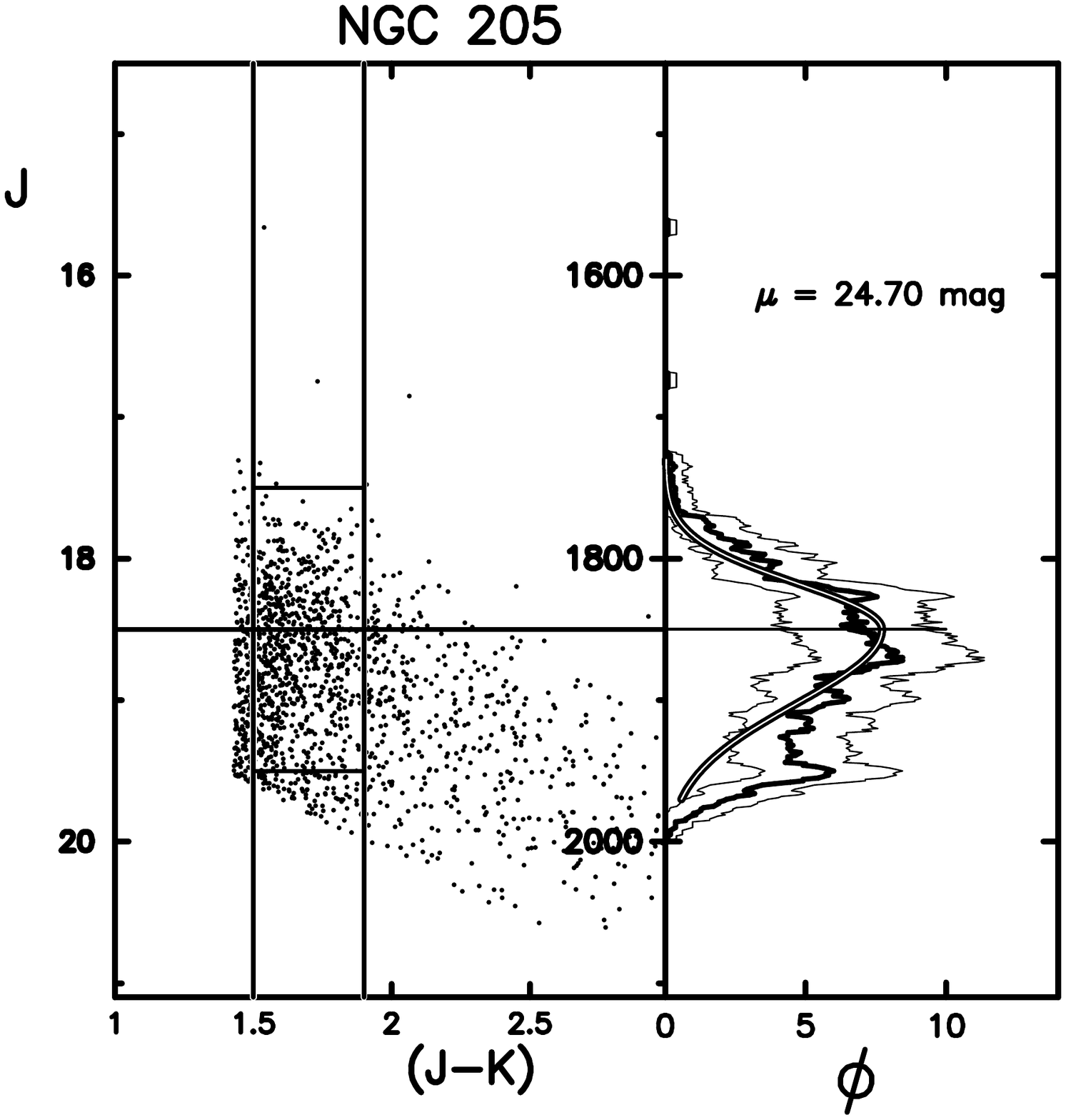} 
\includegraphics[width=5.5cm, angle=-0]{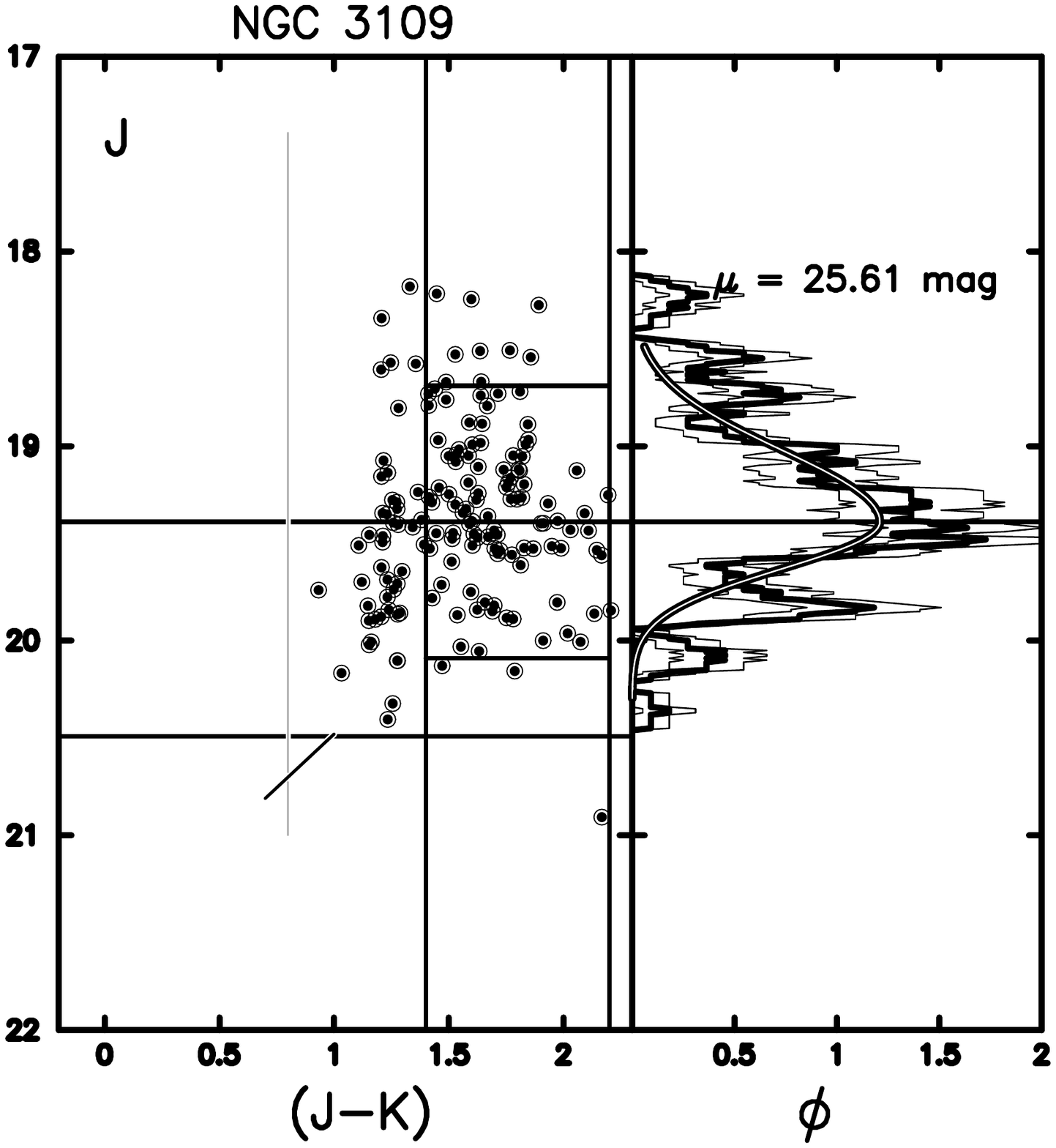} 
\includegraphics[width=5.5cm, angle= 0.0] {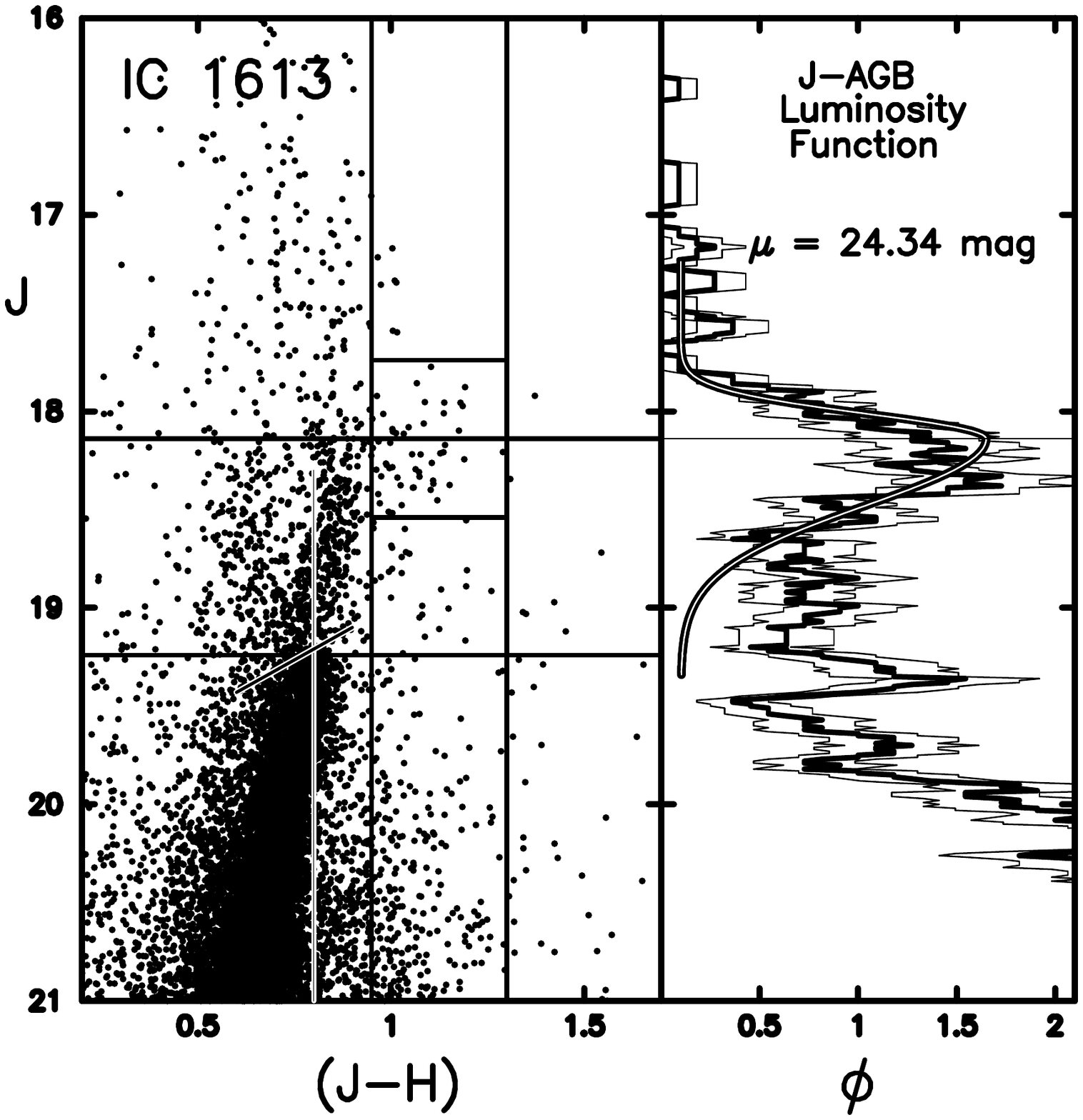}

\includegraphics[width=5.5cm, angle=-0]{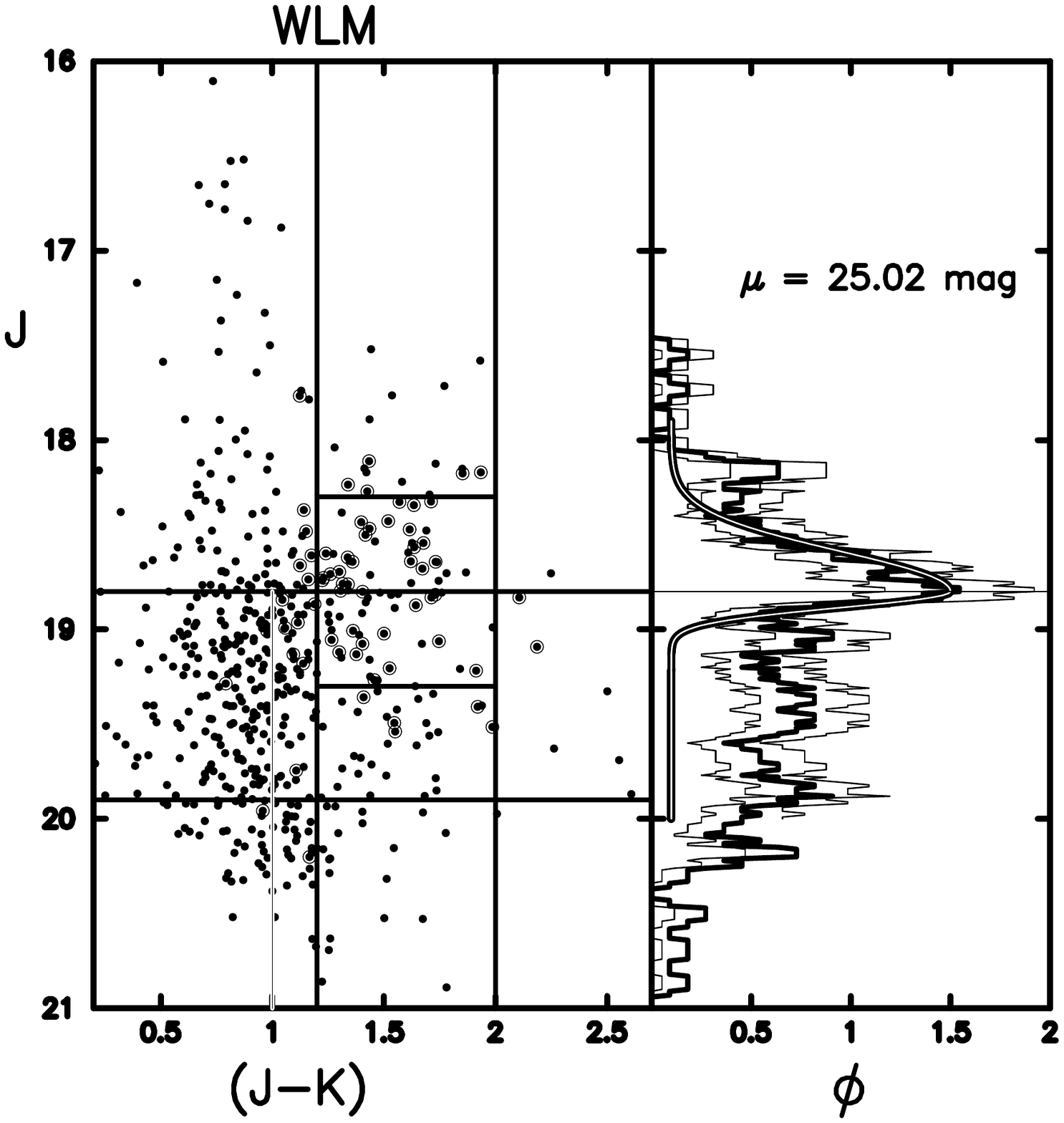}
\includegraphics[width=5.5cm, angle=-0]{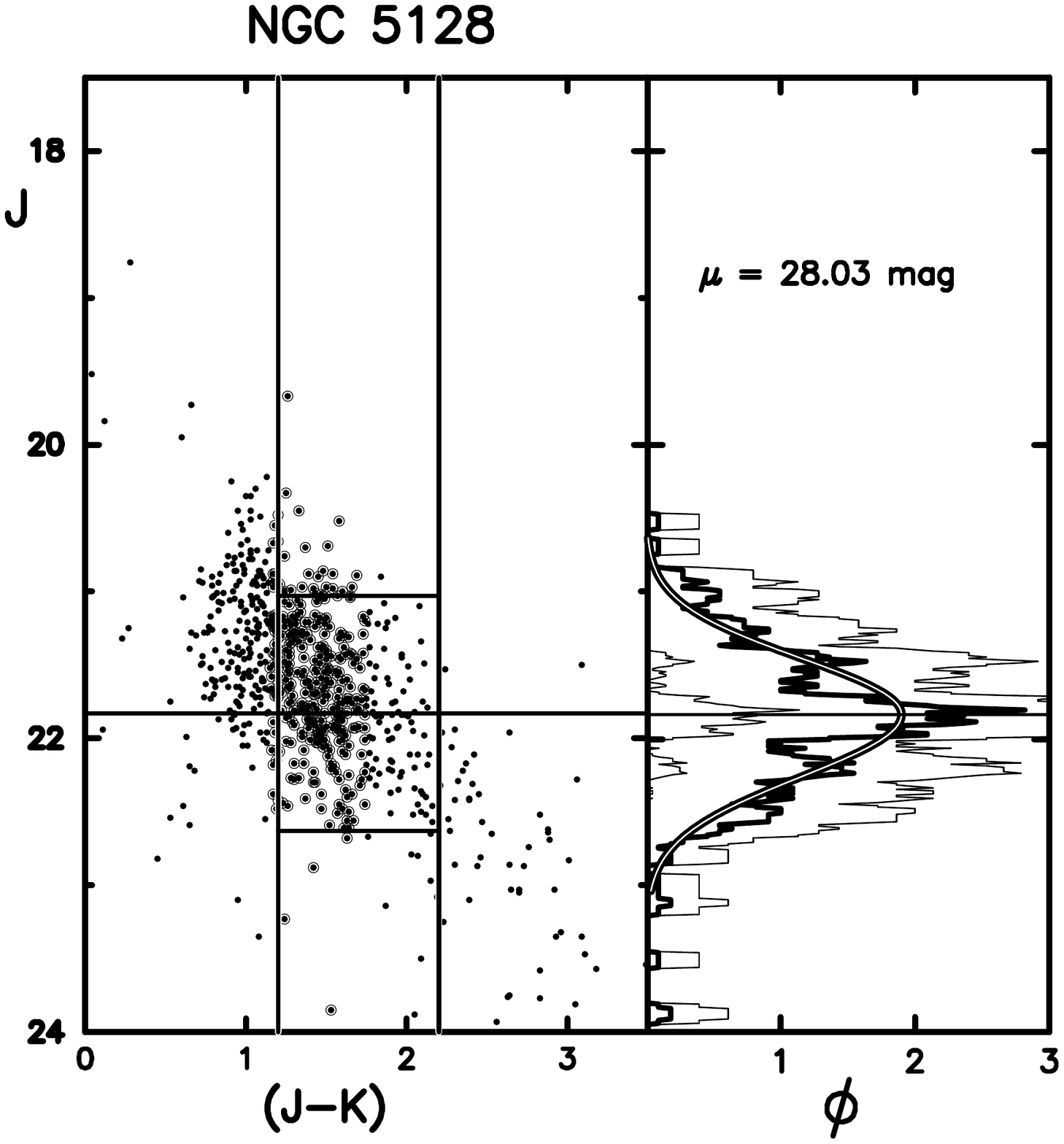}
\includegraphics[width=5.5cm, angle=-0]{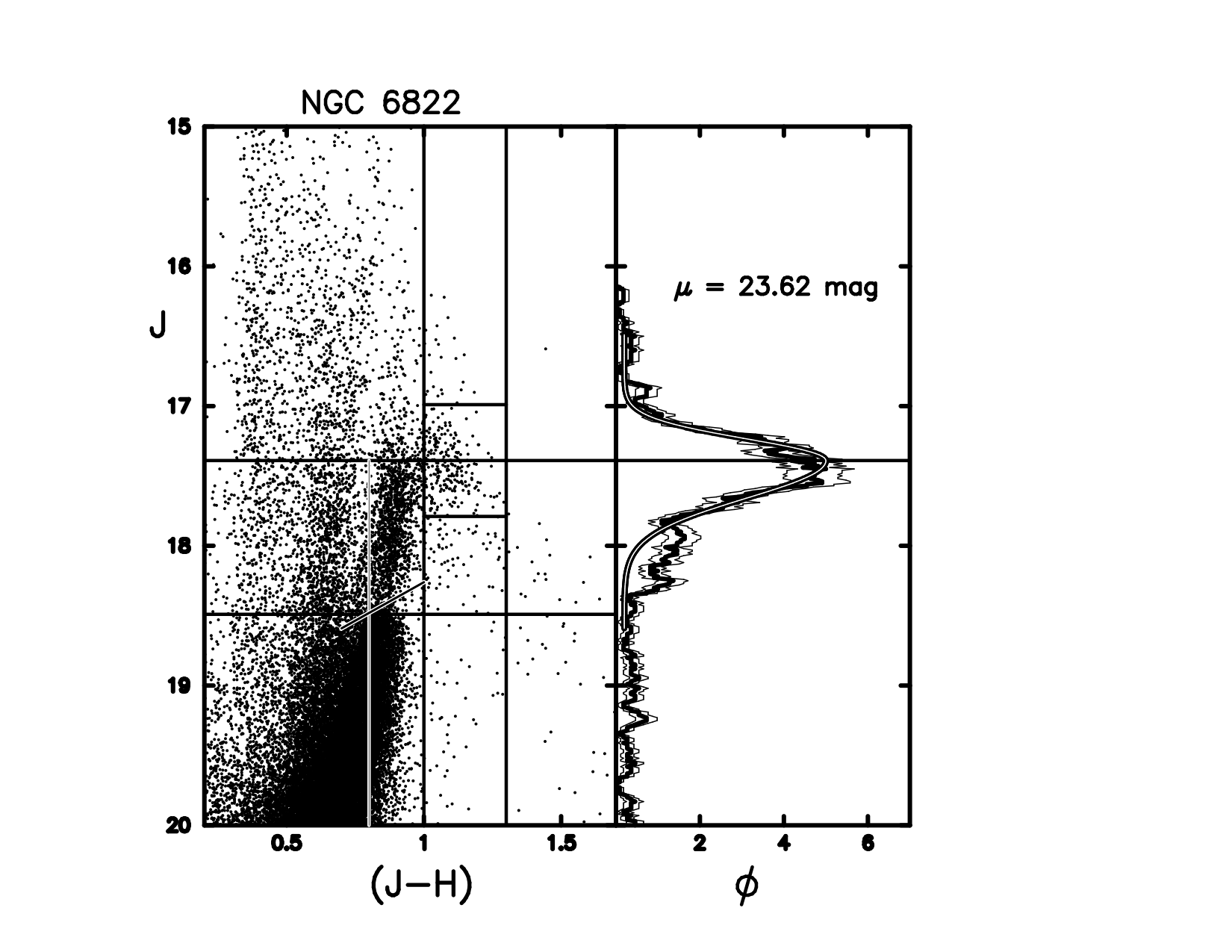} 

\includegraphics[width=5.5cm, angle=-00]{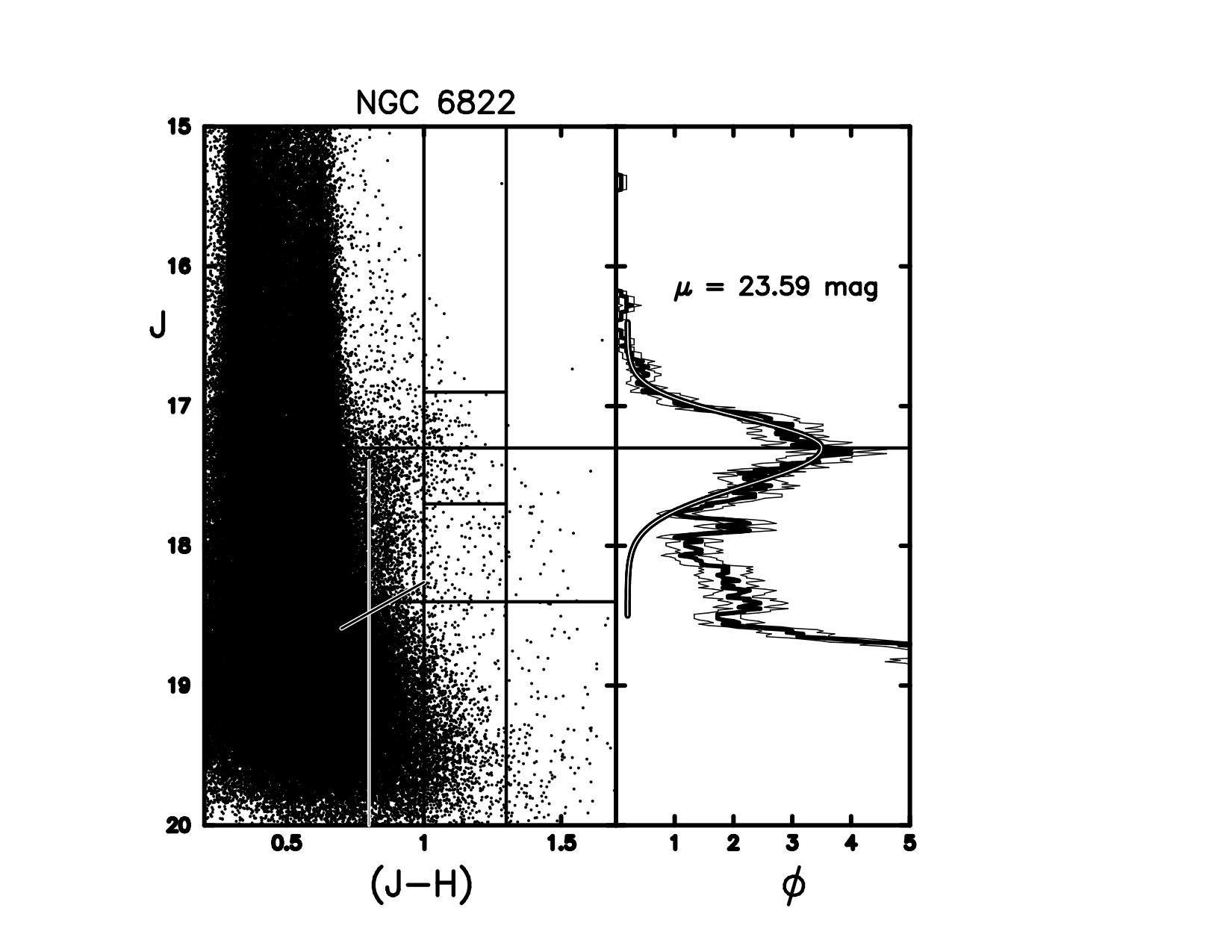} 
\includegraphics[width=5.5cm, angle=-00] {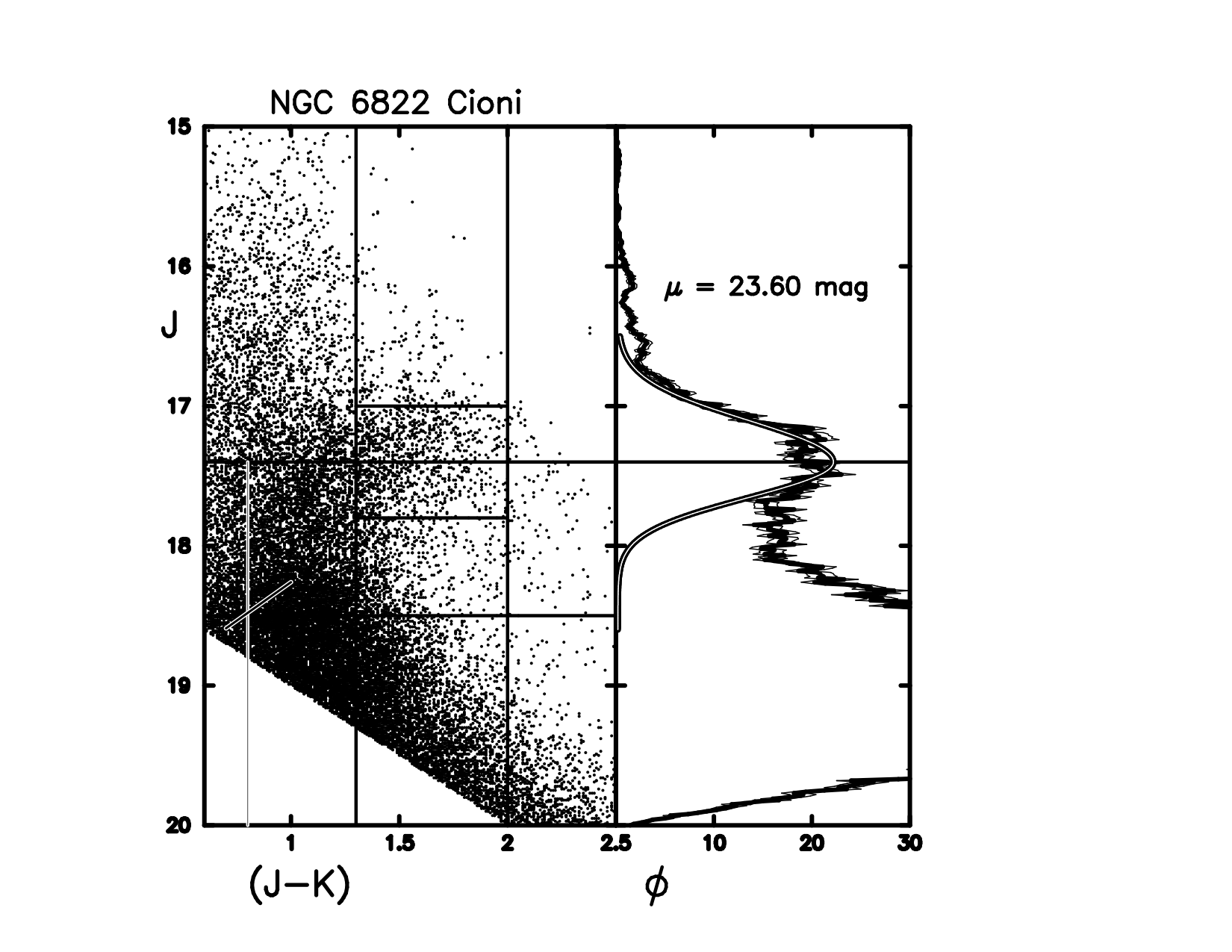} 
\includegraphics[width=5.5cm, angle=-00]{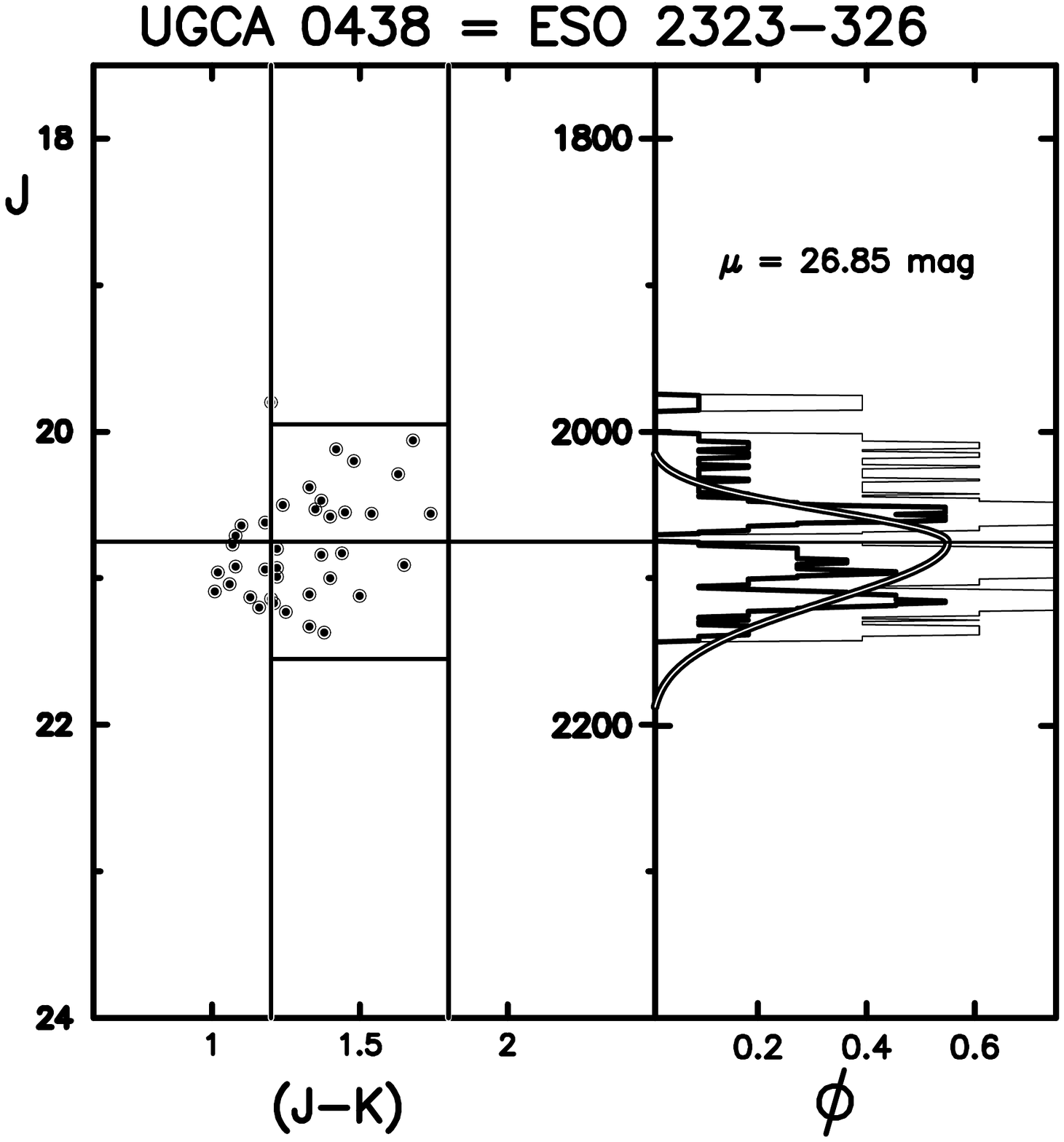}

\includegraphics[width=5.5cm, angle=-00]{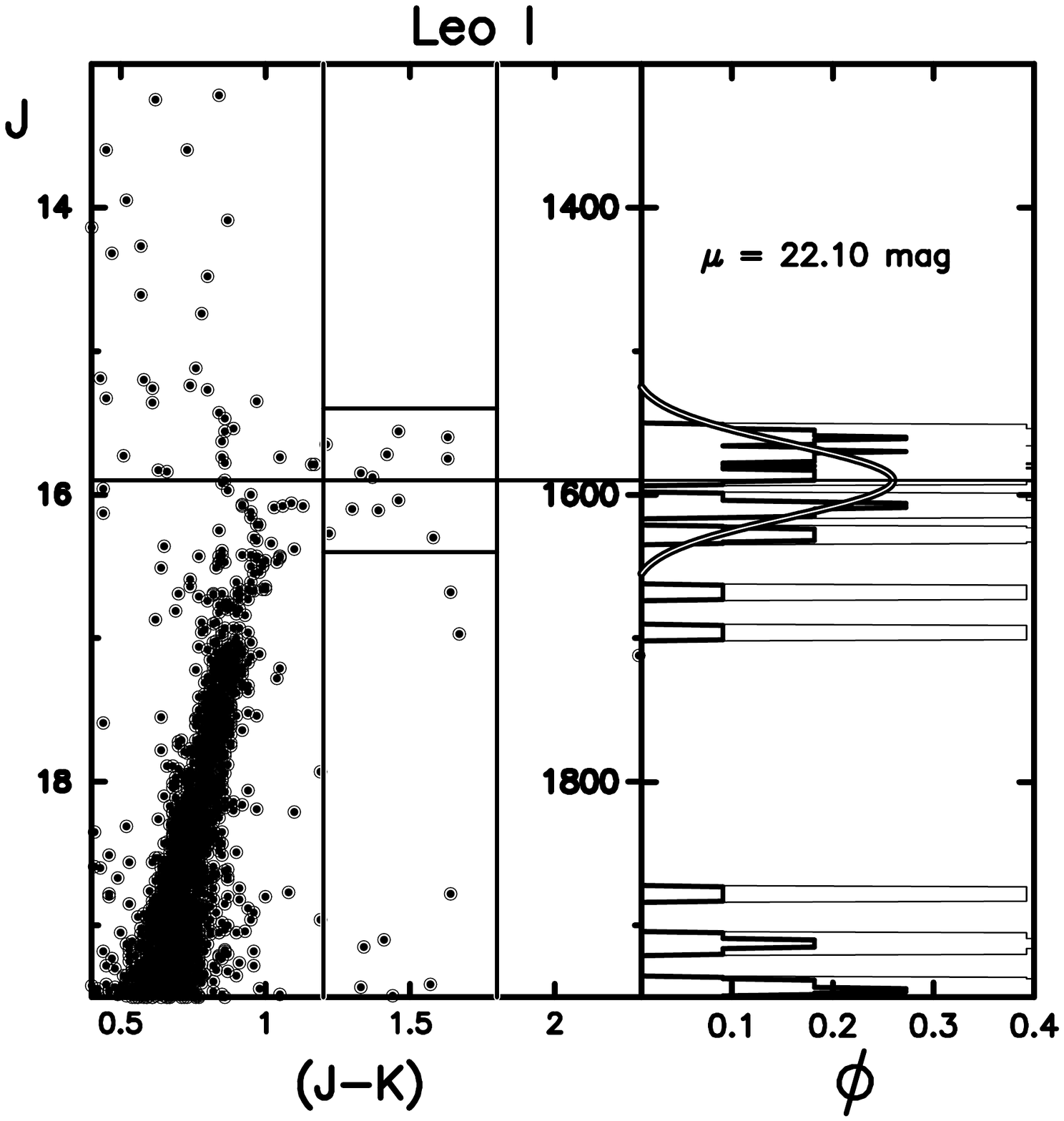} 
\includegraphics[width=5.5cm, angle=-00]{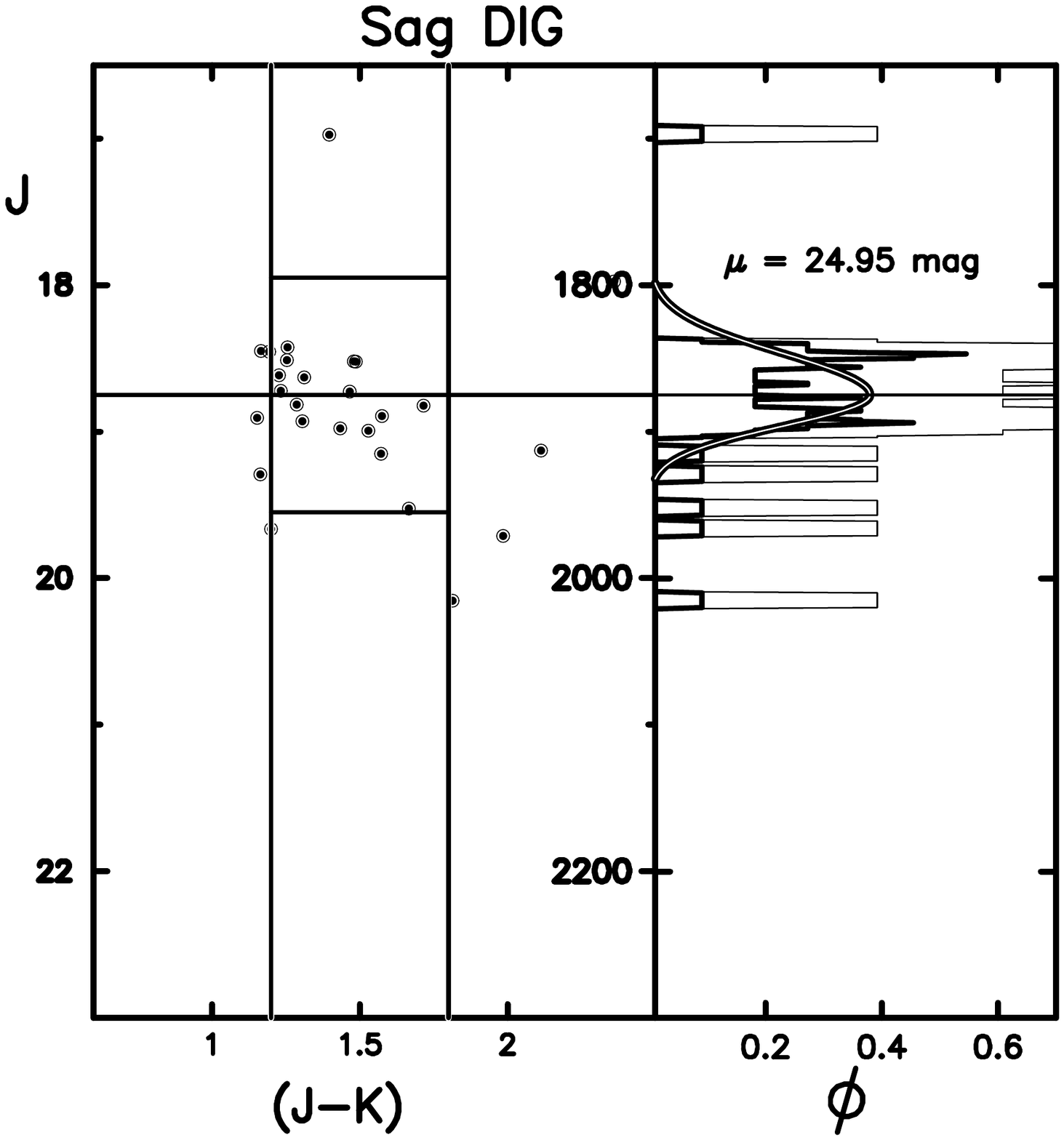}
\includegraphics[width=5.5cm, angle=-00]{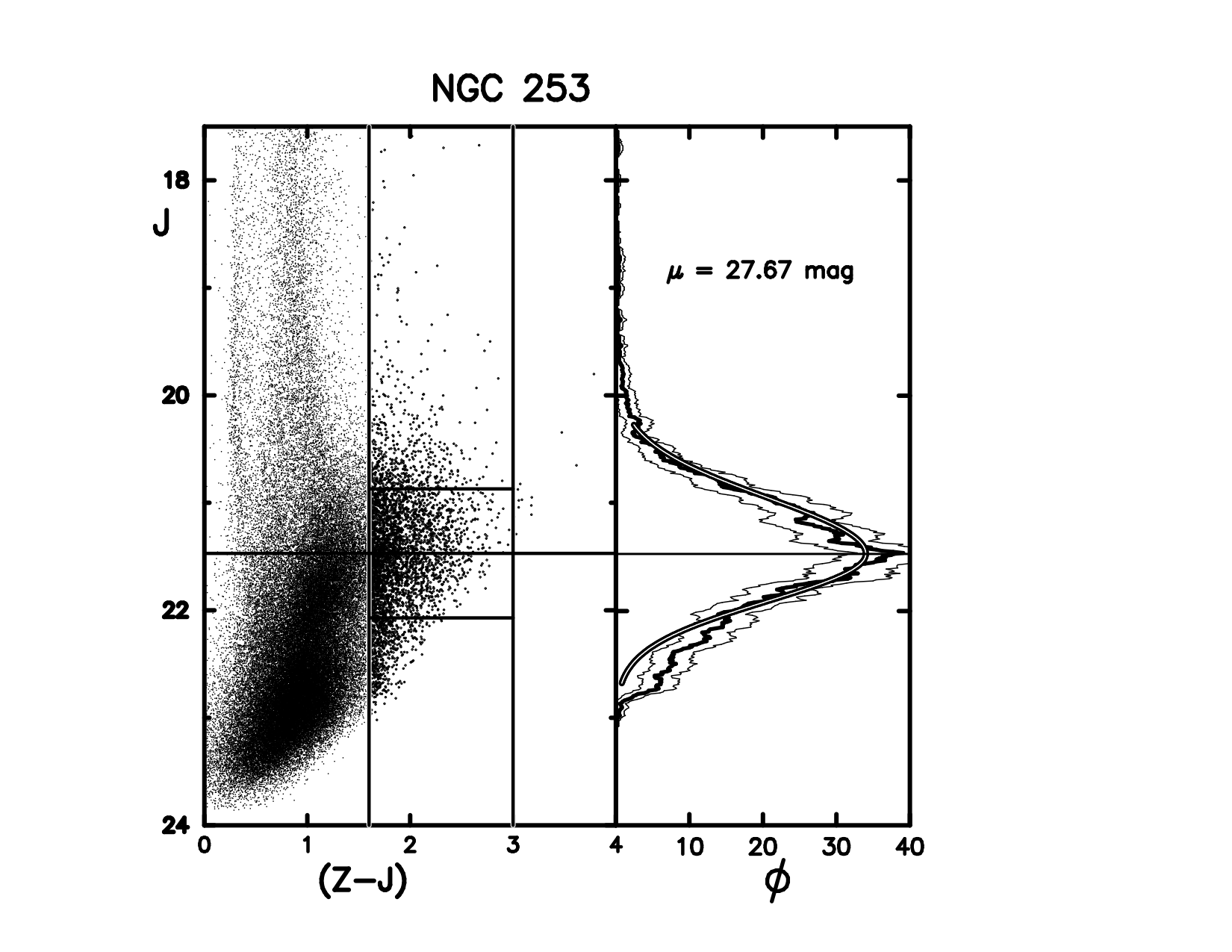}

 \caption{\small Montage of   color-magnitude diagrams (left side of plots) and luminosity functions (right side of plots) for galaxies discussed in this paper; the data for NGC~253 are presented in MF20. The apparent J-band moduli are shown in the individual plots. See the text for details of the individual galaxies. With the exception of Phoenix, CMDs for the galaxies discussed in Appendix \ref{app:othergals} are also displayed here. Both datasets for IC 1613 (discussed in \S\ref{sec:IC1613}) can be found displayed in Figure \ref{fig:IC1613_1}. Plotting details are  the same as Figure \ref{fig: LMC2}. The lines of constant color have been chosen to eliminate contamination by blue oxygen-rich $AGB$ stars and red Region $K$ $AGB$ stars. } 
\label{fig:CMDmontage}
\end{figure}

Each of the galaxies are discussed individually below, in order of increasing distance from the Milky Way. A listing of the newly-derived JAGB distances are given in the second-last column of Table 1, followed by the number of stars in each sample. Median values for the JAGB luminosity are given in brackets in column 6. With the exception of one outlier (IC 1613), the mean difference between the average and the median JAGB luminosity is only 0.003 mag.  Four galaxies, Leo~I, SagDIG, UGCA~438 and Phoenix,  have relatively small samples of measured JAGB stars, while NGC~6822 has a very large foreground extinction, making its distance determination more challenging. These five galaxies are each discussed in Appendix \ref{app:othergals} and are tabulated in Table 1 for completeness. In Figure \ref{fig:CMDmontage} we show the near-infrared $J$ versus $(J-K)$ (or $(J-H)$ or $(Z-J)$) color-magnitude diagrams and the  $J$-band luminosity functions for each of the galaxies. These CMDs and luminosity functions are discussed further below.

\subsection{Fornax Dwarf Spheroidal}

\noindent {\it JAGB stars: } The Fornax dwarf spheroidal galaxy shows no signs of currently on-going star
formation. However, as the first panel in Figure \ref{fig:CMDmontage} shows, Fornax does have an $AGB$ population rising above the TRGB, (seen slightly tilted upwards from about $J = $15.5 to 14.6~mag.) $JHK$ near-infrared photometry for 13,000 stars in Fornax (Gullieuszik et al. 2007a) were used to create the CMD shown in Figure \ref{fig:CMDmontage}. Within the JAGB color-selection box there are 21 luminous, red stars whose luminosity function peaks at $J = $14.58 $\pm$ 0.03~mag. The resulting distance modulus, corrected for a foreground extinction of $A_J = $ 0.015~mag (NED), is $\mu_o ~(JAGB) = $ 20.77 $\pm$ 0.03~(stat)~mag.

\noindent {\it TRGB stars: } There are three unique published TRGB distance moduli for this galaxy. Correcting for a foreground extinction of $A_I$ = 0.032 mag (as given in NED) and adopting $M_I (TRGB) = -4.05$~mag gives: $\mu_o = $ 20.67~mag (Bersier 2000), 20.74~mag (Savianne et al. 2000) and 20.80~mag (Rizzi et al. 2007). These TRGB moduli average out to $<\mu_o (TRGB) > ~= $ 20.73 $\pm$~0.02 ~mag (error on the mean), in good agreement with the estimate using the JAGB method above.

\subsection{NGC 185}
\noindent {\it JAGB stars: } Near-infrared $JHK'$ observations of NGC~185 were obtained by Kang et al. (2005) using the $CFHT-IR$ camera on the CFHT 3.6m at Mauna Kea. Photometry for only the 30 reddest stars was published. The 
resulting CMD is shown in Figure \ref{fig:CMDmontage}. From those data we derive a $J$-star apparent distance modulus 
of 24.14~mag. Applying a Galactic foreground extinction of $A_J = $ 0.130~mag (NED) gives a true 
distance modulus of $\mu_o (JAGB) = $ 24.13~$\pm$~0.06~(stat)~mag.

\noindent {\it TRGB stars: } Five TRGB distances are found in 
the literature.
Corrected for a foreground extinction of $A_I = $ 0.277~mag (NED) and  scaled to $M_I = $ -4.05~mag, the revised true moduli are: $\mu_o = $ 24.00~mag (McConnachie et al. 2005), 24.07~mag (Martinez-Delgado \& Aparicio, 1998), 24.03~mag (Jacobs et al. 2009), 24.07~mag (Salaris \& Cassisi 1997), and 24.08~mag (Butler \& Martinez-Delgado, 2005).  The averaged TRGB true distance modulus is 
then 24.05 $\pm$~0.02~(stat)~mag, a difference of $<$4\% compared to the estimate based on the JAGB method.

\subsection{NGC 147}
\noindent {\it JAGB stars: } Near-infrared $JHK'$ observations of NGC~147 were obtained by Sohn et al. (2006) using the $CFHT-IR$ 
imager on the CFHT 3.6m. Photometry for only the 91 reddest carbon-star candidates was published. The resulting CMD is given in Figure \ref{fig:CMDmontage}. From those data we derive a JAGB apparent distance modulus of 24.53~mag. Applying a Galactic foreground extinction of $A_J = $ 0.122~mag (NED) gives a true distance modulus of $\mu_o (JAGB) = $ 24.41~$\pm$~0.05~(stat)~mag. 

\noindent {\it TRGB stars: } There are four independently-determined TRGB distances published; they are: $\mu_o = $ 24.22~mag (McConnachie et al. 2005), 24.26~mag (Mould \& Sakai 2009), 24.46~mag (Jacobs et al. 2009) and 24.44~mag (Nowotny et al. 2003). The averaged true distance modulus is $<\mu_o (TRGB)> ~= $ 24.35~$\pm$ 0.06~(stat)~mag, agreeing well with the JAGB estimate of the distance to this galaxy.

\subsection{IC~1613}
\label{sec:IC1613}

\begin{figure}[htb!] 
\includegraphics[width=10.0cm, angle=0.0]{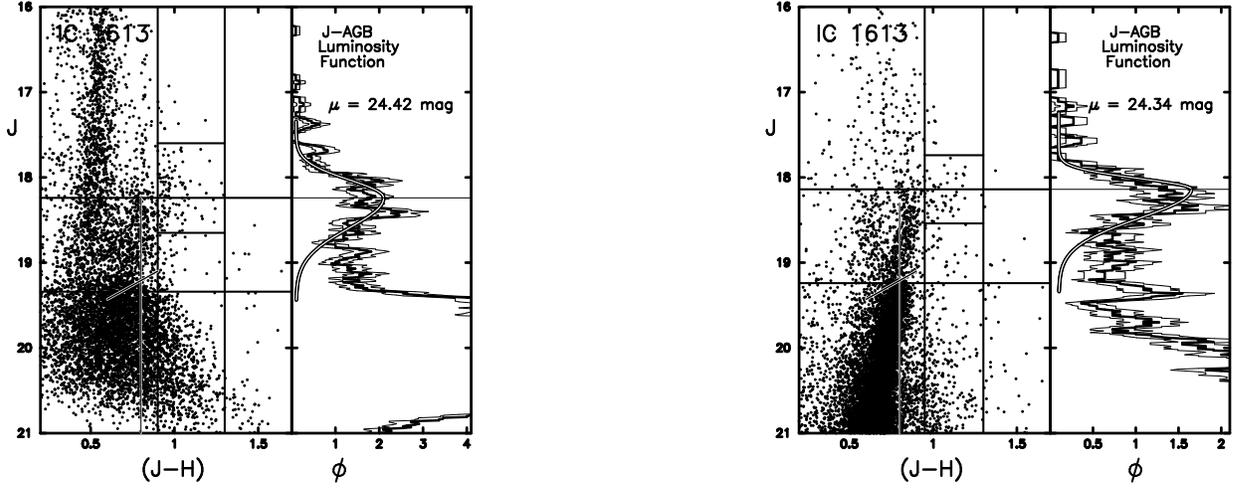} 
\includegraphics[width=10.0cm, angle= 0.0] {IC1613-cmda.pdf}
\caption{\small  Left plot: CMD and luminosity function for luminous red stars in IC~1613 from Sibbons et al. (2015). Right plot: CMD and luminosity function for luminous red stars in IC~1613 from Hatt et al. (2017). Fiducial lines are as described in Figure 2.
} 
\label{fig:IC1613_1}
\end{figure}

\noindent {\it JAGB stars: } For IC~1613 we have two independent data sets for which we can measure a JAGB distance. The first comes 
from a near-infrared ($JHK$) study of the AGB population of IC~1613  undertaken by Sibbons et al. (2015) using the {\it Wide Field Camera} on $UKIRT$, covering 0.8 square degrees on the sky (left-hand side of Figure \ref{fig:IC1613_1}).
The second data set (included in the mosaic Figure 3) comes from Hatt et al. (2017), based on FourStar data taken at the 6.5m Baade Magellan Telescope. In the right-hand panel of Figure \ref{fig:IC1613_1} we show the CMD and luminosity function for IC~1613 from Sibbons et al. (2015) in the same format as given for the LMC. A comparison of the two plots indicates that the Hatt et al. (2017) data are of higher 
signal-to-noise at a given magnitude level and that they go considerably deeper over all, which 
may account for the wider spread seen in the JAGB luminosity function in the Sibbons et al. (2015) plot. The Sibbons et al. field also contains a young, bright and blue supergiant population. 

\noindent The mean of 84 JAGB stars in the Sibbons et al. (2015) study give $J = $ 18.22 $\pm$ 0.01~(stat)~mag.
The Hatt et al. data gives $J = $ 18.14 $\pm$~0.03~(stat)~mag for 41 JAGB stars.
Taking the unweighted mean of these two values, adopting $M_{J_o} = $ -6.20~mag  and applying a foreground extinction correction of $A_J = $~0.018~mag (NED) gives a finally adopted true $J$-band distance modulus of $\mu_o (JAGB) = $ 24.36 $\pm$ 0.03~(stat)~mag.

\noindent {\it TRGB stars: } An overview of recent TRGB distance determinations to IC~1613 by Hatt et al. (2017), rescaled to $M_I = $ -4.05~mag and corrected for a foreground extinction of $A_I = $ 0.038~mag (NED), gives a true distance modulus of $\mu_o (TRGB) = $ 24.36~mag. This is to be compared with $\mu_o (JAGB) = $ 24.36 $\pm$ 0.03~mag found above, as corrected for a foreground extinction of $A_J = $ 0.018~mag (NED).

\subsection{NGC 205}
\noindent {\it JAGB stars: } Jung et al. (2012) 
obtained $JHK$ photometry for 145,246 stars down to a
limiting magnitude of $ K \approx 20.5$ mag; however they
only published a $K \le 18.2$~mag magnitude-selected and
$(J-K) \ge 1.5$~mag color-selected subset of 1,550 red 
AGB stars derived from $WIRCam$ images taken with the CFHT 3.6m. The fall-off of the JAGB luminosity function shown here is not due to incompleteness; their completeness in the $J$ band is estimated to be better than 80\% down to $J =$ 21.5~mag (their Figure 2 top-left panel), which is three magnitudes below the mean $J$ mag of the $JAGB$ population in NGC~205. The resulting $J$ vs $(J-K)$ CMD is shown in Figure \ref{fig:CMDmontage}. Using 876 JAGB stars we derive an apparent $J$-band distance modulus of 24.70~$\pm$~0.01~(stat)~mag. Applying a Galactic foreground extinction of $A_J = $ 0.044~mag (NED) gives a true distance modulus of $\mu_o ~(JAGB) = $ 24.66~$\pm$~0.01~(stat)~mag.

\noindent {\it TRGB stars: } Three published TRGB detections give the following true distance moduli after being  scaled to $M_I = $ -4.05~mag, and corrected for a foreground extinction of $A_I = $0.093~mag: $\mu_o = $ 24.63~mag (Jacobs et al. 2009), 24.61~mag (McConnachie et al. 2005) and 24.76~mag (Butler \& Martinez-Delgado 2005). The averaged true TRGB distance modulus then becomes $<\mu_o ~(TRGB)> = $ 24.67 $\pm$ 0.07~mag.

\subsection{NGC~3109}

\noindent {\it JAGB stars: } Battinelli \& Demers (2009) published $J $ and  $K$ magnitudes for 157 carbon stars in NGC~3109. They are plotted in Figure \ref{fig:CMDmontage}. The luminosity function is found to peak at a mean value of $J = $ 19.41~mag, leading to an apparent distance modulus of 25.61~mag. Correcting for a foreground extinction of $A_J = $ 0.047~mag leads to a true distance modulus of $\mu_o ~(JAGB) = $ 25.56 $\pm$ 0.05~(stat)~mag.

\noindent {\it TRGB stars: } There are nine independent estimates of the TRGB distance to NGC~3109. All consistently corrected for  a foreground extinction of $A_I = $ 0.100~mag (NED) and adopting $M_I(TRGB) = $ -4.05~mag gives $\mu_o = $ 25.50~mag (Lee 1993), 25.44~mag (Dalcanton et al. 2009), 25.55~mag (Mendez et al. 2002),  25.22~mag (Dalcanton et al. 2009),  25.50~mag (Rizzi et al. 2007) 25.69~mag (Hidalgo, Aparicio \& Gallart 2008), 25.65~mag (Minniti, Zijlistra \& Alonso 1999), 25.65~mag (Karachentsev et al. 2002) and 25.62~mag (Jacobs et al. 2009). These values average to $<\mu_o ~(TRGB)> = $ 25.57 $\pm$ 0.05~mag.

\subsection{Wolf-Lundmark-Melotte (WLM)}

\noindent {\it JAGB stars: } Near-infrared ($JK$) data for 555 stars in the distant Local Group galaxy WLM were obtained by Valcheva et al. (2007) using ESO's New Technology Telescope ($NTT$). The $J$ versus $(J-K)$ CMD is shown in Figure \ref{fig:CMDmontage}. Only $JK$ data were available for this galaxy. Circled dots are known carbon stars from Battinelli \& Demers (2004).
Despite the fact that only 77 JAGB stars contribute to the luminosity function, a strong signal is found, peaking at $J$ = 18.80 $\pm$ 0.25~mag, indicating an apparent distance modulus of 25.00 $\pm$ 0.028~mag (stat). Correcting for a foreground extinction component of $A_J = 0.027$~mag (NED) gives a true distance modulus of $\mu_o ~(JAGB) = $  24.97~$\pm$~0.05~(stat)~mag. 

\noindent {\it TRGB stars: } Five TRGB distances have been normalized to $M_I = -4.05$~mag and each corrected for the same adopted 
foreground extinction of $A_I = $ 0.057~mag (NED). The individual true distance moduli are: $\mu_o = $ = 24.79~mag (Minniti \& Zijlstra 1997), 24.84~mag (McCall et al. 2012), 24.85~mag (McConnechie et al. 2005), 24.93 (Rizzi et al. 2007) and 24.95~mag (Jacobs et al. 2009). Their average is $<\mu_o (TRGB)> = $ 24.87 $\pm$ 0.03~(stat)~mag.

\subsection{Cen A = NGC~5128}
\noindent {\it JAGB stars: } In a near-infrared study of Cen A (NGC 5128), Rejkuba et al. (2003) published $JHK$ CMDs for hundreds of long-period variables (LPVs). A sampling of those data are given in Figure \ref{fig:CMDmontage} in which the JAGB stars are prominently visible in the expected color range ($<(J-K)>$~ 1.50 mag). The projection of those stars into their $J$-band luminosity function is shown in the right-hand panel. The apparent distance modulus for Cen~A, based on 162 JAGB stars, is 28.03 $\pm$ 0.032~(stat)~mag. Given a foreground extinction of $A_J$ = 0.081 mag, this reduces to a true distance modulus of $\mu_o ~(JAGB) =~$ 27.95 $\pm$ 0.02 mag.

\noindent {\it TRGB stars: } NED quotes four unique TRGB distance moduli. Uniformly corrected for a foreground extinction of $A_I =$~0.173~mag (NED), the individual true moduli are: 27.78~mag (Lee \& Jang (2016), 
27.98~mag (Hagen et al. 1999), 27.98~mag (Soria et al. 1996) and 27.91~mag (Crnojevic et al. 2013). They average to $\mu~(TRGB) =$ ~27.91 $\pm$ 0.03~(stat)~mag.

\begin{deluxetable*}{lcccccccccr}[b!]
\tablecaption{Intercomparison of TRGB and JAGB  Distance Moduli \label{tab:moduli}}
\tablecolumns{11}
\tablenum{1}
\tablewidth{0pt}
\tablehead{
\colhead{}&\colhead{}&\colhead{TRGB}&\colhead{}&\colhead{}&\colhead{}&\colhead{}&\colhead{JAGB}&\colhead{}&\colhead{} \\
\colhead{Name} &
\colhead{(m-M)$_I$\tablenotemark{a}} & \colhead{$\sigma_I$}&\colhead{A$_I$\tablenotemark{b}} & \colhead{(m-M)$_{Io}$} &\colhead{}&
\colhead{(m-M)$_J$\tablenotemark{f}} & \colhead{$\sigma_J$}& \colhead{A$_J$\tablenotemark{b}} & \colhead{(m-M)$_{Jo}$}&\colhead{No.} \\
\colhead{} & \colhead{mag} & \colhead{mag} & \colhead{mag} &
\colhead{mag } &\colhead{}& \colhead{mag} & \colhead{mag}& \colhead{mag} & \colhead{mag}
&\colhead{JAGB}}
\startdata
Fornax  & 20.76 & 0.02 & 0.032 & {\bf 20.73} && 20.78 (83) & 0.03 & 0.015 & {\bf 20.77} & 21 \\
Leo I\tablenotemark{c}   & 22.10 & 0.04 & 0.055 & {\bf 22.04} && 22.10 (95) & 0.07 & 0.026 & {\bf 22.07} & 12\\
Phoenix\tablenotemark{c} & 23.16 & 0.04 & 0.024 & {\bf 23.14} && 23.16 (16) & 0.18 & 0.014 & {\bf 23.15} &  2\\
NGC 6822\tablenotemark{c} & 23.88 & 0.03 & 0.355 & {\bf 23.52} && 23.64 (66)& 0.02 & 0.167 & {\bf 23.48} &249\\
NGC 0185& 24.33 & 0.02 & 0.277 & {\bf 24.05} && 24.26 (18) & 0.06 & 0.130 & {\bf 24.13} & 30\\
NGC 0147& 29.61 & 0.06 & 0.260 & {\bf 24.35} && 24.53 (50) & 0.05 & 0.122 & {\bf 24.41} & 45\\
IC 1613 & 24.40 & 0.01 & 0.038 & {\bf 24.36} && 24.38 (49) & 0.02 & 0.018 & {\bf 24.36} &125\\
NGC 0205& 24.76 & 0.04 & 0.093 & {\bf 24.67} && 24.70 (76) & 0.01 & 0.044 & {\bf 24.66} &876\\
WLM   & 24.93 & 0.03 & 0.057 & {\bf 24.87} && 25.00 (00) & 0.03 & 0.027 & {\bf 24.97} & 77\\
Sag DIG\tablenotemark{c} & 25.31 & 0.07 & 0.186 & {\bf 25.12} && 24.95 (83) & 0.06 & 0.088 & {\bf 24.86} & 19\\
NGC 3109& 25.67 & 0.05 & 0.100 & {\bf 25.57} && 25.60 (55) & 0.03 & 0.047 & {\bf 25.55} & 83\\
UGCA 438\tablenotemark{c,d} & 26.73 & 0.01 & 0.022 & {\bf 26.71} && 26.85 (82) & 0.06 & 0.010 & {\bf 26.84} & 32\\
NGC 0253\tablenotemark{e} & 27.71 & 0.01 & 0.028 & {\bf 27.68} && 27.67 (69) & 0.01 & 0.013 & {\bf 27.66} &3169\\
Cen A   & 28.08 & 0.03 & 0.173 & {\bf 27.91} && 28.03 (04) & 0.03 & 0.081 & {\bf 27.95} &162\\
\enddata
\tablenotetext{a}{~Uniformly adopting $M_I$ = -4.05~mag and individual NED foreground reddenings}
\tablenotetext{b}{~NED Release Jan. 2020; based on Schlafly \& Finkbeiner (2011)}
\tablenotetext{c}{~Discussed in Appendix \ref{app:othergals}}
\tablenotetext{d}{~UGCA~438 = ESO 2323-326}
\tablenotetext{e}{Discussed in Madore \& Freedman (2020)}
\tablenotetext{f}{The quantity in brackets is the decimal portion of the apparent distance modulus derived from the median.} 

\end{deluxetable*}

\section{Comparisons with TRGB Distances}
\label{sec:compare}

\begin{figure}[htb!] 
\centering
\includegraphics[width=12.5cm, angle=-0]
{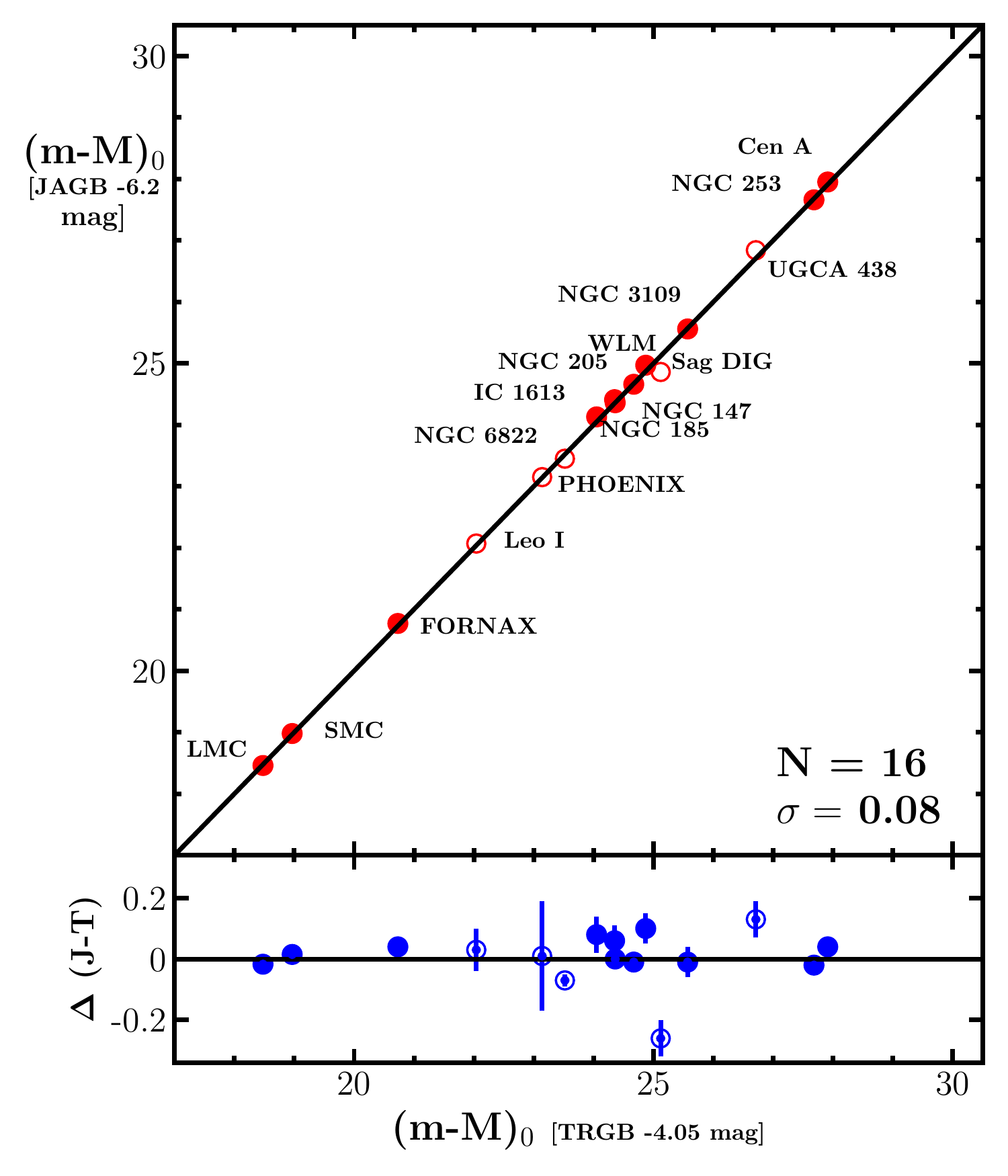} \caption{\small Comparison of 16 {\it JAGB} distance moduli with the
TRGB distance moduli cited in the text.  A unit-slope line is shown, but is not a fit to the points. Points with open circles designate galaxies for which there are currently small numbers of stars observed, therefore having lower statistical weight.} 
\label{fig:TJComp}
\end{figure}

In Figure \ref{fig:TJComp}, we show a comparison of TRGB and JAGB distances
for the 13 galaxies discussed in this paper, along with data for NGC ~253, the LMC and the SMC from MF20. The  16 galaxies in Figure \ref{fig:TJComp} have a scatter
about a unit-slope line of $\pm$0.08~mag. Excluding the largest outlier (SagDIG), the mean zero-point 
offset, in the sense (JAGB) minus (TRGB) $=$ +0.025 $\pm$ 0.013~mag (scatter on the mean), or a 1.2\% agreement in the two distance scales. (Including SagDIG, the difference is even smaller, +0.007 $\pm$ 0.022 mag, although both estimates are consistent to within their given uncertainties.)
{\t That is, the independently adopted
zero points of the TRGB and JAGB calibrations are in good agreement.}
If the scatter seen in the lower panel of Figure \ref{fig:TJComp} 
(i.e., $\pm$0.08~mag) is shared equally between the two methods, then 
the intrinsic scatter independently attributable to each of the methods comes to
$\pm$0.06~mag. This small scatter suggests that, in principle, each 
of these two methods can individually provide  distances that are
statistically good to 3\% or better. (Alternatively, if the TRGB turns out to be a more precise  distance
indicator, so that it contributes insignificantly to the combined scatter, then this comparison still suggests that JAGB distances would still
individually be good to better than 4\%.) 

Having found that the two (TRGB and JAGB)
distance scales are in extremely good statistical agreement, we further note that changing the distance modulus to the LMC, upon which both methods depend for their respective zero points, will not change that differential agreement. In addition, this current agreement provides independent support for the adopted TRGB absolute magnitude of $M_I = $~-4.05~mag  (Freedman et al. 2020).

Finally, we point out that the observed scatter in the JAGB distance moduli calculated with respect to the TRGB method must include the cumulative effects of all of the differences  resulting from the host galaxy types, metallicities, star-formation histories and uncorrected differential reddening effects in the AGB populations. That is, all systematics contained in this sample, and all random differences between these galaxies, in excess of photometric and fitting errors (which are conservatively estimated here to be at the $\pm$0.05 mag level in each of the two methods), are themselves constrained at the $\pm$0.05 mag level or less.


\section{Potential Sources of Uncertainty}
\label{sec:errors}

\subsection{Reddening}
\label{sec:e_red}

We now turn to the issue of  corrections for line-of-sight reddening to the JAGB stars in our sample of galaxies. The Galactic (Milky Way) foreground contribution can be estimated in a straightforward way from the maps derived by  Schlafly \& Finkbeiner (2011) and provided on-line on a galaxy-by-galaxy basis by NED, as included in Table 1, and described in \S4. The contribution from reddening internal to the host galaxy is more challenging, and will be discussed in more detail in a future paper, as we continue to refine and improve the technique. There are at least three possibilities for dealing with internal extinction. The first is to use a reddening-free Wesenheit function (Madore 1982), as used for the Cepheid Leavitt law.   The second is to use some of the well-defined vertical sequences, as already identified by WN01, to act as fiducial ``lines'' from which to derive reddenings. The oxygen-rich $AGB$ population (Region \emph{F}) is one possibility, rising as it does nearly vertically at  $(J-K)_o \sim $ 1.2~mag. An alternate approach to dealing with reddening is to move the observations further to the red than the currently adopted $J$-band.  For example, if the uncertainty in adopting a host-galaxy reddening is $\epsilon_{E(B-V)} = \pm$ ~0.040~mag then $\epsilon_{A_J} = \pm$~0.032~mag, but $\epsilon_{A_K} = \pm $~0.020~mag, or 1\% in the distance error budget.

In addition to the foreground reddening component internal to the galaxy,
there is also the possibility that JAGB stars have a component of
self-extinction from dust created in their atmospheres. In this exploratory
stage in the calibration, we  provisionally assume that the latter two
reddening components are, on average, the same in 
the calibrator (LMC/SMC) samples as they are in the other galaxies (see also
the discussion by Huang et al. (2020) in their study of Mira variables in
NGC ~1559).  Positive or negative differences in this case will simply
average out; i.e., they would not contribute a systematic offset. However,
while still developing this technique and its calibration, we have 
provisionally adopted an uncertainty in the visual ($A_V$) extinction of
$\pm$0.10~mag, which translates into a $J$-band systematic (extinction) error of $\pm$0.03~mag.

We note that the JAGB luminosity function being discussed in this paper excludes the extreme carbon-star (X-carbon star) population, which is bright in the mid infrared, and known to be affected by dust obscuration (e.g.,  Boyer et al. 2011). As mentioned above, these stars have been labeled as Region K stars in WN01 and Figure \ref{fig:LMC}.

We have begun a pilot program to obtain high-precision $JHK$ data for nearby galaxies, from which we can measure reddening-free Wesenheit, $W$, magnitudes. In a preliminary study of the galaxy M33 (in preparation), for a sample of 2,450 JAGB stars, we find that the scatter in $W$ is $\pm$0.29 mag, smaller than the scatter in $J$-band of $\pm$0.33 mag (as would be expected if $W$ were correcting for reddening). In the case of the LMC, WN01 also quote a scatter of $\pm$0.33 mag for the $K$-band relation. This test suggests that with high-precision photometry, as for nearby Cepheids (e.g., Freedman et al. 2001), the use of the Wesenheit function  will be an important tool in further reducing systematic uncertainties in the application of the JAGB method.

\subsection{Star Formation History and Metallicity}

\subsubsection{Star Formation History}
\label{sec:e_sfh}There is a long history in the literature providing evidence that the total carbon star luminosity function (CSLF) (e.g., Marigo et al. 1999) and  carbon star magnitudes correlate with star formation history (Frogel et al. 1990; Searle, Wilkinson \& Bagnuolo 1980). The carbon to M star (C/M) ratio is also found to depend on age of the population (e.g., Cioni et al. 2006, Pastorelli et al. 2019 and references therein). In principle, the star formation history of a galaxy could have an effect on the JAGB peak magnitude.  

In the context of distance scale determination, it is important to distinguish between the total CSLF and that of the JAGB LF. The total CSLF is measured using carbon stars identified spectroscopically or with intermediate-band filters. It is known to include both bluer  and redder  stars than those contained in the JAGB LF (defined by 1.4 $< (J-K) <$ 2.0~mag).  The J Region stars are identified on the basis of their near-infrared colors alone. Battinelli \& Demers (2007) showed that very few M stars have $(J-K) >$ 1.4~mag, and hence do not contaminate the J region.  (The total CSLF however,  includes a non-negligible contribution of carbon stars with bluer colors than 1.4~mag, which cannot be distinguished from M stars on the basis of  $(J-K)$ colors alone.) The ratio of C/M stars also often includes the extreme carbon-star (X-carbon star) population with $(J-K)$ colors redder than 2.0~mag, the so-called Region K stars in Weinberg \& Nikolaev (2001). As such, they do not contribute to the JAGB luminosity function being discussed here.

In summary, the JAGB LF turns out empirically to have a stable peak in its luminosity function with a very small scatter. However, this is not expected to, and does not apply to the total CSLF in general. Both the earlier study of WN01 and the current study suggest that the JAGB method provides an excellent standard candle, which is amenable to further empirical testing. In addition, the near-constant luminosity and small observed dispersion in this subset of carbon stars offers theorists a feature that can provide an added constraint on models of AGB evolution.

\vfill\eject
\subsubsection{The C/M Ratio}
\label{sec:e_cm}

Many studies have found a correlation between [Fe/H] metallicity and the ratio of C to M stars in nearby galaxies (e.g., Cook et al. 1986; Battinelli \& Demers 2005; Cioni \& Habing 2003, 2005; Cioni 2009; Boyer et al. 2019). The correlation is in the sense that the C/M ratio decreases with increasing metallicity. Theoretical models suggest that the correlation is driven primarily by two effects: the smaller number of M stars in metal-poor systems and the fact that for lower metallicity, the oxygen-rich stars do not require as much carbon to be dredged up  to result in  a carbon star with C/O $>$ 1 (e.g., Iben \& Renzini 1983). Models also suggest that the C/M ratio depends on the star formation history of the galaxy (e.g., Cioni et al. 2006; Marigo et al. 2013), in addition to the helium abundance (Karakas 2014). 

As we noted in Section \ref{sec:compare} above, the distances determined from the JAGB LF show only a modest (4\%) scatter in comparison to independently measured TRGB distance moduli. In addition, the ability of WN01 to measure the tilt of the LMC, suggests that metallicity (and age) effects cannot dominate the signal of J-region luminosity function. In the following subsection, we address the question of whether empirically, the JAGB peak luminosity itself shows a correlation with metallicity.

\subsubsection{A Test for a Metallicity Effect}
\label{sec:e_met}

\begin{figure}[htb!] 
 \centering
\includegraphics[width=8.5cm, angle=-0]
{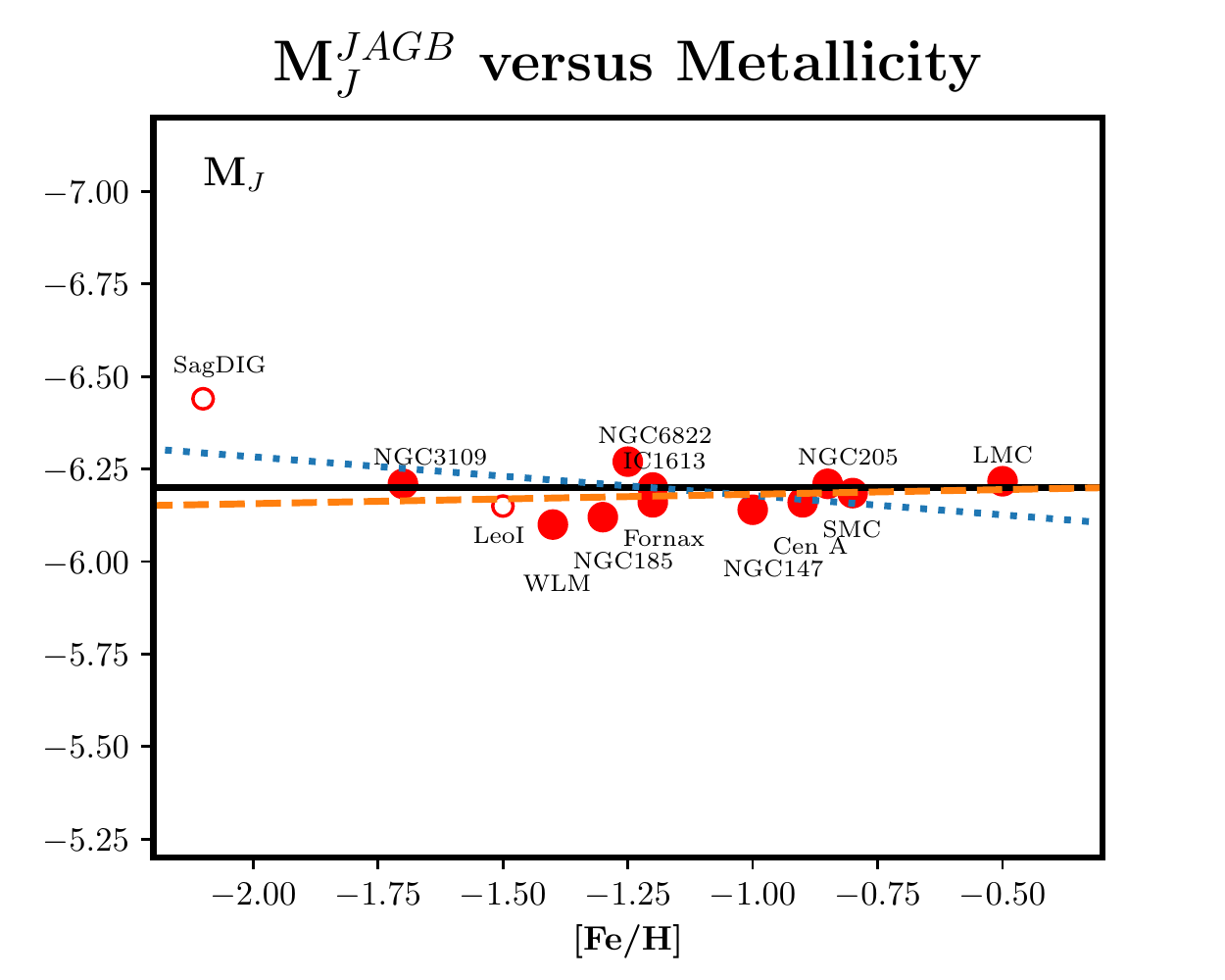} \caption{\small The absolute $J$-magnitude, $M_J^{JAGB}$ for JAGB stars as a function [Fe/H] metallicity of the parent galaxy. The [Fe/H] values shown as red filled circles are from  Table 1 of Battinelli \& Demers (2005), except for Cen A, for which the metallicity is from Bird et al. (2015). The black horizontal line is the adopted zero point derived from DEB distances to the LMC and SMC, set to M$_J$ = -6.2~mag. The  blue dotted line includes all galaxies for which Battinelli \& Demers tabulated metallicities; the orange dashed line is a least-squares fit excluding SagDIG and Leo I, which have very few JAGB stars. For this sample of galaxies measured in the  $J$-band, the sensitivity to metallicity of the parent galaxy is not statistically significant.  } 
\label{fig:metals}
\end{figure}

In an initial attempt to investigate the possibility of second parameters in the J-band luminosity of JAGB stars, Figure \ref{fig:metals} shows the residuals from Figure \ref{fig:TJComp} plotted as a function of host galaxy [Fe/H] metallicity, as tabulated by Battinelli \& Demers (2005). The dotted blue line represents an unweighted fit to the current sample of 12 galaxies with metallicity measurements from the Battinelli \& Demers compilation. The slope of this line is  +0.10 $\pm$ 0.06 mag/dex; a marginally significant result, largely dependent upon one galaxy, SagDIG, found at the very low-metallicity extreme. The dashed orange line represents a second unweighted fit, this time eliminating the two galaxies shown as open circles: SagDIG and Leo I, for which there are only 19 and 12 JAGB stars, respectively, defining their distances. The slope of this fit is  -0.03 $\pm$ 0.05 mag/dex. Without these two low-weight galaxies, the resulting slope is closer to zero (with even lower statistical significance), and its sign has reversed. At this early level of scrutiny, metallicity apparently does not play a measurable role in modulating the absolute $J$-band magnitude of the JAGB.  This constraint can be improved in future, as more galaxies have JAGB distances measured. This near-infrared comparison can be contrasted with that from Battinelli \& Demers (2005; their Figure 4) based on $I$-band photometry, for which the slope of the metallicity relation appears to be steeper than observed at $J$.

\subsection{Mass Loss}
\label{sec:e_massloss}

Mass loss remains a difficult challenge for models, governing both AGB evolution and termination (e.g., Willson 2000; Höfner \& Olofson 2018; Pastorelli et al. 2019 and references therein). The modeling of thermally pulsating AGB stars depends on the understanding of a number complex processes, many of which are interconnected: for example, third dredge up, hot bottom burning, pulsations, circumstellar dust, and dust-driven winds. 

Recently, Pastorelli et al. (2019) have used the CSLF to calibrate how metallicity affects the dredge-up process as well as mass loss in AGB stars. These models are based on near- and mid-infrared observations of AGB stars in the SMC and LMC, which suggest that the highly evolved, extreme X-AGB stars, while a small fraction of the total number of AGB stars ($\sim$5\%), dominate the dust contribution by about an order of magnitude (Boyer et al. 2015; Srinivasan et al. 2016).   

As noted in Section 6.2.1, the X-AGB (Region K) stars are  eliminated from the JAGB sample. The issue for use of the JAGB method in future applications, is to ascertain how the peak of the JAGB luminosity function might be affected by circumstellar, dust-driven mass loss. Our early test using the Wesenheit function (Section 6.1) suggest that reddening-free magnitudes will be an important and effective tool for minimizing the effects of dust.

\subsection{Variability of JAGB Stars}
\label{sec:e_var}

As noted by WN01, most of the stars in Region J are C-rich thermally pulsing AGB stars. Based on the study of Wood (2000), most of these stars are semi-regular variables, with some Miras. The fraction of all stars in this region that are variable at some level, is close to 100\%. If unaccounted for, the presence of variability will increase the scatter in the JAGB LF, but not contribute a systematic uncertainty. 

WN01 measured the width of the main peak in their $K$-band LF in a narrowly-defined range of ($J$-$K$) color), and found that for the stars in Region J, $\sigma \approx$~0.2 mag. (As we note, using the $J$- (rather than $K$-) band LF, which is flat with color, there is no need to make such a narrow color selection; Madore \& Freedman 2020). WN01 note that this scatter includes not only the intrinsic precision of the Region J stars as a standard candle, but, in addition, the variance due to differential extinction, the geometric depth of the LMC, and photometric errors in the photometry. Thus this scatter represents an {\it upper limit} to the intrinsic scatter of the peak of the JAGB LF, and suggests that these stars are indeed excellent standard candles. 

\subsection{Foreground Contamination}
\label{sec:e_foregd}

Could foreground stars affect the result, particularly in low latitude galaxies? Nikolaev \& Weinberg (2000) (their Section 3) describe the contributions of Galactic foreground stars to the regions defined by them in the LMC. The contribution by foreground stars occurs primarily at $(J-K)$ $<$ 1.0~mag (their Regions B and C), but Region J  (at the redder colors $(J-K)$~$>$~1.4~mag) is comprised primarily of thermally pulsing carbon stars.

Further evidence that foreground stars are not a significant contribution can be seen in a comparison of the CMDs for NGC 6822 and IC 1613. NGC 6822 lies at a low galactic latitude of -18 deg, and can be compared with IC 1613, which is out of the galactic plane at a latitude of -61 deg.  From Figure \ref{fig:CMDmontage}, it can be seen  that contamination by foreground stars at these red $(J-K)$ colors of the JAGB appears to be minimal. 

In future, observations with a built-in cadence could help to identify not only JAGB stars exhibiting variability, but also non-variable objects (including Galactic foreground stars, as well as unresolved background galaxies), and thereby minimize contamination of the JAGB population.

\subsection{Discussion}
\label{sec:e_discuss}

Concluding this section, we emphasize that our application of the JAGB method is based on a specific subset of near-infrared, photometrically-selected carbon stars, the J-region as defined by Weinberg \& Nikolaev (2001). This subset of stars appears (empirically) to show little, if any, dependence on metallicity or age of the population. 

\noindent
There are two direct pieces of evidence supporting this statement: 

\noindent
1) Distances based on the luminosity of the peak of the JAGB LF agree to within a combined scatter  +/- 0.08 mag (4\%) of those based on the (independent) TRGB distances (\S\ref{sec:compare}), and 
\par\noindent
2) WN01 used the $K$-band JAGB LF (for a limited color range) to measure the geometric tilt of the LMC, a measurement in excellent agreement with previous authors’ estimates for the tilt. If age and metallicity and circumstellar dust effects were a large or dominant effect on the luminosity of these stars, for example, no correlation with position could have been detected. And, as WN01 note, the scatter in their observed luminosity function must also include contributions from differential reddening, line of sight depth effects, and photometric errors, in addition to the intrinsic scatter in the luminosity function; and yet the total observed scatter is still very small.

\section{Comparison of Mira Variables and JAGB Stars as Distance Indicators}

In this section we compare the relative merits of the period-luminosity relations for Mira variables as compared to the JAGB method, and their practical application to the extragalactic distance scale. We begin with the Mira variables.
\label{sec:Miras}

\subsection{Mira Variables}

Mira variables have been used to measure the distance to the Large Magellanic Cloud for many years; in particular, based on detailed work by Feast and collaborators (Feast et al. 1989; Feast 2004; Whitelock et al. 2008). These long-period variables have been divided into two classes based on their spectral features: oxygen-rich and carbon-rich.   Oxygen-rich (O-rich) Miras have a carbon-to-oxygen C/O ratio $<$ 1 and  carbon-rich (C-rich) Miras have C/O $>$ 1. All high-luminosity AGB stars begin oxygen-rich (with M-type spectra), but as discussed in \S\ref{sec:theory} carbon can, in certain instances, be `dredged up' into the atmosphere from the interior, leading to the formation of carbon-rich atmospheres (and thus C-type spectra). In the optical, the amplitudes of Miras are large, amounting to  $>$ 2.5 mag in the $V$-band, but falling to  $>$ 0.4 mag in the $K$-band. In the $K$-band, the scatter in the period-luminosity relation for O-rich Miras amounts to only $\pm$0.14 mag (according to Whitelock et al. 2008). 

A recent, detailed and comprehensive discussion of the issues faced in applying Mira variables to the extragalactic distance scale beyond the LMC is given in Huang et al. (2018). The main advantage of long-period variables, in general,  is that they are both numerous and bright; moreover they can be found in galaxies of all morphological types. The challenge, however, is that LPVs, as a class, are a heterogeneous set of stars that range from irregular, to semi-regular, to periodic variables, with stars pulsating in their fundamental and higher overtone modes, while still unpredictably fluctuating in amplitude, phase and period. Without spectra, it is a non-trivial endeavor to identify which stars belong in which of the several competing period-luminosity relations seen at the same luminosity level.

Below we list these and many of the other challenges facing the establishment of a Mira variable distance scale:

(a)	In order to discover Mira variables and measure their periods, phases, amplitudes and mean magnitudes, the observing windows must be at least as wide as the longest-period variable being studied, so that, at very least, one complete cycle is covered. Known Mira variables have periods ranging from 100 up to 1,000 days or more. Determining periods then requires multiple observations over a very long time baseline. Even if the targeted sample is narrowed to periods less than 400 days, say, the discovery of Mira variables becomes a very telescope-time intensive endeavor, requiring observational campaigns spanning several years, at least.

(b)	One pulsation cycle may not be enough, given that Miras exhibit cycle-to-cycle variability (e.g., see Mattei 1997 for a review). Accordingly, for better accuracy in measuring their mean photometric properties, sampling two or more cycles is required to provide greater confidence in their measured periods at a minimum. Moreover, the regular Mira variables have magnitudes and colors that are similar to irregular and semi-regular, long-period-variable $AGB$ stars, from which they must be somehow distinguished and extracted. A graphic example of this complexity of AGB variability can be seen in Figure 1 of Soszyński et al. (2009), which shows the (overlapping) period-luminosity relations for the many different kinds of long-period variables, including OGLE Small Amplitude Red Giants (OSARGs), oxygen-rich Miras, carbon-rich Miras and semi-regular variables (SRVs). To complicate things further, there is a mix of overtone pulsators. 

(c) Selecting high-confidence Mira variables, of the O-rich sub-type,requires high-precision periods, amplitudes and additionally, multi-wavelength color information (see Table 3 of Huang et al. 2018). The noise from the dominant AGB population of LPVs, the instability of the best Mira light curves, and the huge temporal baselines required to detect and measure the light-curve/periodic properties of these stars, each pose significant challenges for this implementing this technique, especially when spectroscopic data are unavailable.

\subsection{JAGB Stars}

In stark contrast with having to deal with the systematics and complexities of sorting through a mixed variable star population to obtain distances, the simplicity of the JAGB technique provides a major advantage over Miras in their application to the extragalactic distance scale. For JAGB stars only a single-epoch, near-IR CMD is required to isolate and measure the peak of their near-infrared luminosity function, and then straight-forwardly derive distances. No further down-selecting within the AGB sub-populations is required. No time domain observations, or periods, amplitudes, light curves or time-averaging are needed.  The JAGB distance determination is based on the population sample mean, whose intrinsic dispersion is known to be $\pm$0.2 mag in the near-infrared $J$ band (WN01, and as illustrated in Figure 2 for the LMC.) However, with 2-4 epochs, mean magnitudes can be obtained for the JAGB stars, averaged over any and all types of variability (or not), and thereby readily decreasing the dispersion in the observed luminosity function. In all, the JAGB method is about an order of magnitude more efficient in observing time than that required for finding the O-rich Mira variables.

Since LPV variability is intrinsic to many (or most) JAGB stars, this means that simply averaging two or more, randomly-observed epochs will bring down the observed scatter (as 1/$\sqrt{N_e}$, where $N_e$ is the number of widely-spaced epochs). Averaging can be applied without needing to have any knowledge of the light curves, periods, phases or amplitudes: one simply can split the exposure times and treat the underlying variability simply as another source of noise. 

A clear example of the low yield and overall complexity of attempting to use the period-luminosity relation for oxygen-rich Miras as a distance indicator is illustrated by the case of M33. Based on a decade of observations, Yuan et al. (2017a,b, 2018) identified 239,907 variables in M33, and obtained light curves for them during their extended search for Mira variables. This sample was then culled down to 1,847 Mira candidates. But, ultimately only 853 stars that had near-infrared photometry passed a 3-$\sigma$ rejection criterion, giving a 0.4\% yield on the original search. Had there been only 1,000 variable stars discovered, only 4 Miras would have been left after the down-selcetion process.  On the other hand, with a single-epoch visit, JAGB stars can be identified using a simple color selection criterion and have their distances determined from a simply-computed mean magnitude of the ensemble; moreover, the JAGB approach is robust to any underlying variability of a sub-population of those extremely red stars, periodic or not.

We conclude that the JAGB method offers numerous and significant advantages (of telescope-time and simplicity) over the use of the oxygen-rich, Mira-selected (amplitude-selected, color-selected and sigma-clipped) period-luminosity relation. 
\section{Looking to the Future}

In this section we discuss several topics that need to be addressed as we move forward with the understanding, calibration and application of the JAGB method.

First and foremost is the task of producing a homogeneous survey of the nearest galaxies using standard $JK_s$ filters, and acquiring complete samples of stars at high signal-to-noise, sampling a range of star formation histories and metallicities.

Second, beyond correcting for Milky Way line-of-sight extinction, we need to explore the reddening of the JAGB population within the host galaxies, as well as dealing with extinction that may be affecting the JAGB stars locally (e.g., {\it in situ} formation of dust.)  This will involve exploring reddening-free (Wesenheit) magnitudes in the near infrared as has been successfully developed for Cepheids. High-precision data in at least two bands will be needed.

Finally, if this method is to be pushed to its logical limit in its application 
to the extragalactic distance scale and to an independent determination of the
Hubble constant, then it is paramount that a secure tie-in is constructed between the ground-based photometric system being used for the nearest galaxies and the
flight-magnitude near-infrared systems on board HST, JWST and WFIRST.

\section{Summary}

Here we summarize the numerous attributes of JAGB stars as a class of easily-discovered and simply-measured, high-luminosity stars, showing their potential as  high-precision and high-accuracy extragalactic distance indicators. This discussion augments that given earlier by Battinelli \& Demers (2005), now emphasizing the developing application of this technique at near- and mid-infrared wavelengths for space-based determinations of the extragalactic distance scale
using \hst, \jwst and WFIRST.
\par\noindent
(1) JAGB stars are distinctive, and the dominant population of the reddest stars found in galaxies. They are, therefore,
easily identified on the basis of their near-infrared colors, without the need for spectroscopy (or narrow-band
filters). Since spectroscopic classification of AGB stars is impractical for most, if not all extragalactic purposes,
color selection is the only plausible  way forward, if a significant number of galaxies are to be surveyed. 
\par\noindent
(2) After a simple color selection, the $J$-band luminosity function is
unambiguously defined, centrally peaked, and has low dispersion. Calibrated in the LMC and SMC, the absolute $J$-band magnitude of the JAGB sequence is $M_J =$ ~-6.20 $\pm~0.20/\sqrt{N_c}$ mag (where $N_c$ is the number of JAGB stars in the marginalized luminosity function). 
\par\noindent
(3) In the mean, JAGB stars are about one magnitude brighter than the brightest Population II stars, i.e., those defining the tip of the red giant branch (TRGB).
\par\noindent
(4) JAGB stars are found in all galaxies that have intermediate-age populations. The JAGB method is, therefore, applicable to a wide range of galaxy types, from spirals and irregulars, to dwarf spheroidals (e.g., Fornax), and even dwarf ellipticals (such as NGC~185 and NGC~205, discussed in Section 4 above).

\par\noindent
(5) Infrared observations of JAGB stars offer at least two major advantages over the optical: 
\par (a) The peak-to-peak amplitude of light variability within the population is significantly reduced in going from the optical to the infrared, dropping, in some cases, by as much as a factor of 10.
\par (b) The total line-of-sight reddening (composed of Milky Way foreground extinction, host galaxy extinction, and the possibility of locally-confined dust reddening generated by the individual carbon stars themselves) drops by a factor of nine in going from the optical $V$-band, for example, to the near infrared $K$-band. For example, if $A_V =$ ~0.20 mag of extinction were left unaccounted for in the optical, this would give rise to a systematic error of 10\% in the distance. However, this would reduce to only $A_K$~= 0.022 mag in the K-band, resulting in a systematic error in the distance of only 1\%.
\par\noindent
(6) Unlike Cepheids or Mira variables, which must be discovered, typed and followed over multiple cycles for their variability, the JAGB population of stars is simply color-selected. No variability criteria are required for JAGB stars to be used as distance indicators. However, some form of variability is common, if not universal, among the stars making up this JAGB population (including LPVs, Miras and semi-regular variables).  Single-epoch (i.e., random-phase) observations of these variables will add to the intrinsic width of the luminosity function; but, without requiring any knowledge of the phases, amplitudes or periods of the variability, this form of “photometric noise” can be averaged down with one or more additional random-epoch observations, which can be taken months or years later. No periods and no light-curve fitting are required; just a small number of random-phase epochs.
\par\noindent
(7)  Reddening within the host galaxy is currently the largest potential systematic uncertainty for the JAGB population. However, this effect can be minimized by using a reddening-free Wesenheit magnitude.

Given this extensive list of positive attributes, we are optimistic 
that the JAGB method will be competitive in future studies of the
distances to individual nearby galaxies, and for independently extending the extragalactic distance scale. We re-emphasize that the publicly available samples of JAGB stars (and their photometry) analyzed in this study were not optimized in any way for the application of this technique. With further targeted observations, an increase in the sample sizes, and careful attention to building a consistent photometric calibration, the possibility for significantly increasing the accuracy of this technique, beyond what has been done in this first application, is extremely promising. Currently our test sample extends out to a distance of $\sim$4~Mpc. The reach of the  JAGB method is considerably greater, both with \hst, and especially in the near future when \jwst becomes operational.  

In summary, we find excellent agreement between  previously-published TRGB distance estimates and the JAGB determinations.  The  16 galaxies in our sample have a scatter about a unit-slope line of only $\pm$0.08~mag. Excluding the largest outlier (SagDIG), the mean zero-point offset ($JAGB - TRGB$) $=$ +0.025 $\pm$ 0.013 ~mag. 


\section{Acknowledgements}
This research made daily use of the NASA/IPAC Extragalactic Database (NED), 
which is operated by the Jet Propulsion Laboratory, California Institute 
of Technology, under contract with the National Aeronautics and Space 
Administration. We thank the {\it Observatories of the Carnegie Institution for
Science} and the {\it University of Chicago} for their support of our long-term research
into the calibration and determination of the expansion rate of the Universe. Financial 
support for this work was provided in part by NASA through grant number HST-GO-13691.003-A 
from the Space Telescope Science Institute, which is operated by AURA, Inc., under 
NASA contract NAS 5-26555.  We thank Elias Oakes and Alex Masegian for their help in preparing figures. 
In addition, we thank all of the authors, cited in the text, who made their data publicly 
available for easy access by the rest of the community. Finally, our thanks to an anonymous referee for very constructive suggestions. 

\section{References}

\medskip

\noindent
Aubourg, {\'E}., Bailey, S., Bautista, J.~E., et al.\ 2015, \prd, 92, 123516

\noindent
Battaglia, G., Rejkuba, M., Tolstoy, E., et al. 2012, \mnras, 424, 1113

\noindent
Battinelli, P. \& Demers, S. 2004, A\&A, 416, 111

\noindent
Battinelli, P. \& Demers, S. 2005, A\&A, 442, 159

\noindent
Battinelli, P., Demers, S., \& Mannucci, F.\ 2007, \aap, 474, 35

\noindent
Battinelli, P. \& Demers, S. 2009, A\&A, 493, 1075

\noindent
Beccari, G., Bellazzini, M., Fraternali, F., et al. 2014, A\&A, 507, 78

\noindent
Bellazzini, M., Gennari, N., Ferraro, F.~R. \& Sollima, A. 2004, \mnras, 354, 708.

\noindent
Bersier, D. 2000, ApJ, 543, 23

\noindent
Bird, S.A., Flynn, C., Harris, W.E., et al. 2015, A\&A, 575, 72

\noindent
Boothroyd, A.~I., Sackmann, I.-J., \& Ahern, S.~C.\ 1993, \apj, 416, 762

\noindent
Boyer, M.~L., Srinivasan, S., van Loon, J.~T., et al.\ 2011, \aj, 142, 103

\noindent
Boyer, M.~L., McQuinn, K.~B.~W., Barmby, P., et al.\ 2015, \apj, 800, 51

\noindent
Boyer, M.~L., Williams, B.~F., Aringer, B., et al.\ 2019, \apj, 879, 109

\noindent
Butler, D.~J., \& Martinez-Delgado, D. 2005, AJ, 129, 2217

\noindent
Cabrera-Lavers, A. \&  Garzon, F. 2004, AJ, 127, 1386

\noindent
Caputo, F., Cassisi, S., Castellani, M., et al. 1999, AJ, 117, 2199

\noindent
Carpenter, J. 2003, http://astro.caltech.edu/$\sim$jmc/2mass/v3/transformations

\noindent 
Cioni, M.-R.~L.\ 2009, \aap, 506, 1137  

\noindent
Cioni, M.-R.~L., Girardi, L., Marigo, P., et al.\ 2006, \aap, 448, 77

\noindent
Cioni, M.-R.~L., \& Habing, H.~J.\ 2003, \aap, 402, 133

\noindent
Cioni, M.-R.~L. \& Habing, H.~J. 2005, A\&A, 429, 837

\noindent
Cook, K.~H., Aaronson, M. \& Norris, J. 1986, \apj, 305, 634

\noindent
Crnojevic, D., Ferguson, A.M.M., Irwin, M.J. et al. 2013, \mnras, 432, 832 

\noindent
Dalcanton, J.~J., Williams, B.~F., Seth, A.C. et al. 
2009, ApJS, 183, 67

\noindent
Feast, M.~W., Glass, I.~S., Whitelock, P.~A., et al. 1989, \mnras, 241, 375

\noindent
Feast, M. 2004, ASPC, 310, 304

\noindent
Feeney, S.~M., Mortlock, D.~J., \& Dalmasso, N. 2018, \mnras, 476, 3861

\noindent
Freedman, W.~L. 2017, NatAs, 1, 121

\noindent
Freedman, W.~L., Madore, B.~F., Gibson, B.~K., et al.\ 2001, \apj, 553, 47

\noindent
Freedman, W.~L., Madore, B.~F., Scowcroft, V., et al. 2012, \apj, 758, 24

\noindent
Freedman, W.~L., Madore, B.~F., Hatt, D. et al. 2019 \apj, 882, 34

\noindent
Freedman, W.~L., Madore, B.~F., Hoyt, T. et al. 2020, \apj, 891, 57

\noindent
Frogel, J.~A., Mould, J., \& Blanco, V.~M.\ 1990, \apj, 352, 96

\noindent
Fusco, F., Buonanno, R., Bono, G., et al. 2012, A\&A, 548, 129

\noindent
Gonzales-Fernandez, C., Hodgkin, S.T., Irwin, M.J., et al. 2015, \mnras, 472, 5459

\noindent
Gullieuszik, M., Held, E.~V., Rizzi, L., et al. 2007a, A\&A, 467, 1025 

\noindent
Gullieuszik, M., Rejkuba, M., Cioni, M.-R., Habing, H.~J. \& Held, E.~V. 2007b, A\&A, 475, 467  

\noindent
Gullieuszik, M., Greggio, L., Held, E.V., et al. 2008, A\&A, 483, L5

\noindent
Habing, H.~J. \& Olofsson, H. 2004, ``Asymptotic Giant Branch Stars'',  Springer-Verlag, New York

\noindent
Hagen, G.L.H., Harris, W.E., \& Poole, G.B. 1999, AJ, 117,855

\noindent
Hatt, D., Beaton, R.~L., Freedman, W.~L., et al. 2017, \apj, 845, 146

\noindent
Held, E.~V., Savianne, I., \& Momany, Y. 1999, A\&A, 345, 747

\noindent
Held, E.~V., Gullieuszik, M., Rizzi, L., Girardi, L, Marigo, P \& Saviane, I. 2010, MNRAS, 404, 1475 

\noindent
Herwig, F.\ 2013, Planets, Stars and Stellar Systems. Volume 4: Stellar Structure and Evolution, 397

\noindent
Hidalgo, S.~L., Aparicio, A., \& Gallart, C.\ 2008, \aj, 136, 2332

\noindent
Hidalgo, S.L., Aparicio, A., Martinez-Delagada, D. \& Gallart, C. 2009, \apj, 705, 704

\noindent
H{\"o}fner, S., \& Olofsson, H.\ 2018, \aapr, 26, 1

\noindent
Huang, C.~D., Riess, A.~G., Hoffmann, S.~L., et al. 2018, \apj, 857, 67

\noindent
Huang, C.~D., Riess, A.~G., Yuan, W., et al.\ 2020, \apj, 889, 5

\noindent
Iben, I. \& Renzini, A. 1983, ARAA, 21, 271

\noindent
Jang, I.~S., \& Lee, M.~G.\ 2017, \apj, 836, 74

\noindent
Jung, M.~Y., Ko, J., Kim, J.-W. 2012, A\&A, 543, 35

\noindent
Jacobs, B.~A., Rizzi, L., Tully, R.~B., et al. 2009, AJ, 138, 332

\noindent
Kang, A., Sohn, Y.-J., Kim, Ho-II., Rhee, J., Kim, J.-W., Hwang, N., Lee, M.-G., Kim, Y.-C., \& Chun, M.-S. 2006, A\&A, 454, 717 

\noindent
Karachentsev, I.~D., Aparicio, A., \& Makarov, D.~E. 1999, A\&A, 352, 363

\noindent
Karachentsev, I.~D., Sharina, M.~E., Makarov, D.~E., et al. 2002, A\&A, 389, 812

\noindent
Karakas, A.~I.\ 2014, \mnras, 445, 347

\noindent
Kochanek, C.~S.\ 2020, \mnras, 493, 1725

\noindent
Lee, M.-G. 1993, \apj, 408, 409

\noindent
Lee, M.-G. \& Byun, Y.-I. 1999, AJ, 118, 817

\noindent
Lee, M.~G., Freedman, W.~L., \& Madore, B.~F.\ 1993, \apj, 417, 553

\noindent
Lee, M.-G. \& Jang, I.-S. 2016, \apj, 822, 70

\noindent
Macaulay, E., Nichol, R.~C., Bacon, D., et al.\ 2019, \mnras, 486, 2184

\noindent
Madore, B.~F.\ 1982, \apj, 253, 575

\noindent
Madore, B.~F. \& Freedman, W.~L. 2020, \apj, submitted

\noindent
Macri, L., Ngeow, C.-C., Kanbur, S., Mahzooni, S., \& Smitka, M.~T. 2015, AJ, 149, 117

\noindent
Marigo, P., Bressan, A., Nanni, A., et al.\ 2013, \mnras, 434, 488

\noindent
Marigo, P., Girardi, L., \& Bressan, A.\ 1999, \aap, 344, 123

\noindent
Marigo, P., Girardi, L., Bressan, A., et al. 2008, A\&A, 482, 883

\noindent
Marigo, P., Girardi, L., Bressan, A., et al. 2017, \apj, 835, 77

\noindent
Martinez-Delgado, D. \& Aparicio, A. 1998, AJ, 115, 1462

\noindent
Martinez-Delgado, D., Gallart, C. \& Aparicio, A. 1999, AJ, 118, 862

\noindent
Mattei, J.~A. 1997, JAVSO, 25, 57

\noindent
McCall, M.~L., Vaduvescu, O., Pozo Nunez, F., et al. 2012, A\&A, 540, 49

\noindent
McConnachie, A.~W., Irwin, M.~J., Ferguson, A.~M.~N., et al. 2005, \mnras, 356, 979 

\noindent
Mendez, B., Davis, M., Moustaka, J., et al. 2002, AJ, 124, 213

\noindent
Menzies, J., Feast, M., Whitelock, P., Olivier, E., Matsunaga, N. \& da Costa, G. 2008, \mnras, 385, 1045 

\noindent
Minniti, D. \& Zijlstra, A.~A. 1997, AJ, 114, 147

\noindent
Minniti, D., Zijlistra, A~.A. \& Alonso, M.~V. 1999, AJ, 117, 881

\noindent
Momany, Y., Held, E.~V., Saviane, I., \&  Rizzi, L. 2002, A\&A, 384, 393

\noindent
Mould, J.~R. \& Sakai, S. 2009, \apj, 694, 1331 

\noindent
Nikolaev, S. \& Weinberg, M.~D., 2000, \apj, 542, 804

\noindent
Nowotny, W., Kerschboum, F., Olofsson, H. \& Schwarz, H.~E. 2003, A\&A, 403, 93  

\noindent
Pastorelli, G., Marigo, P., Girardi, L., et al.\ 2019, \mnras, 485, 5666

\noindent
Pietrzynski, G., Graczyk, D., Gieren, W., et al. 2010, AJ, 140, 1038

\noindent
Pietrzynski, G., Graczyk, D., Gallenne, A., et al. 2019, Nature, 567, 200

\noindent
Planck Collaboration, Ade, P.~A.~R., Aghanim, N., et al.\ 2016, \aap, 594, A13

\noindent
Planck Collaboration, Aghanim, N., Akrami, Y., et al.\ 2018, eprint, arXiv:1807.06209

\noindent
Pritchet, C.~J., Richer, H.~B., Schade, D. et al. 1987, \apj, 323, 79

\noindent
Riess, A.~G., Macri, L.~M., Hoffmann, S.~L., et al.\ 2016, \apj, 826, 56
 
\noindent
Riess, A.~G., Casertano, S., Yuan, W., et al. 2019, \apj, 876, 85

\noindent
Rich, J., Persson, S.~E., Freedman, W.~L., Madore, B.~F., et al. 2014, \apj, 794, 107

\noindent
Richer, H.~B, 1981, \apj, 243, 744

\noindent
Richer, H.~B. \& Crabtree, D.~R. 1985, \apj, 298, 240

\noindent
Richer, H.~B., Crabtree, D.~R \& Pritchet, C.~J. 1984, \apj, 287, 138

\noindent
Richer, H.~B., Pritchet, C.~J. \& Crabtree, D.~R. 1985, \apj, 298, 240

\noindent
Rizzi, L., Tully, R.~B., Makarova, D., et al.  2007, \apj, 661, 815

\noindent
Rosenfield, P., Marigo, P., Girardi, L., et al.\ 2014, \apj, 790, 22

\noindent
Salaris, M. \& Cassisi, S. 1997, \mnras, 289, 406

\noindent
Salaris, M., Weiss, A., Cassar{\`a}, L.~P., et al.\ 2014, \aap, 565, A9

\noindent
Savianne, I., Held, E.~V. \& Bertelli, G. 2000, A\&A, 355, 56

\noindent
Schlafly, E.~E. \& Finkbeiner, D.~P. 2011, ApJ, 737, 103 

\noindent

\noindent
Searle, L., Wilkinson, A., \& Bagnuolo, W.~G.\ 1980, \apj, 239, 803

\noindent
Sibbons, L.~F., Ryan, S.~G., Cioni, M.-R.~L., Irwin, M. \& Napiwtzki, R. 2012, A\&A, 540, 135 

\noindent
Sibbons, L.~F., Ryan, S.~G., Irwin, M. \& Napiwtzki, R. 2015, A\&A, 573, 84 

\noindent
Sohn, Y.-J., Kang, A., Rhee, J., Shin, M., Chun, M.~S., \& Kim, Ho-II. 2006, A\&A, 445, 6 

\noindent
Soria, R., Mould, J.R., Waston, A.M. et al. 1996, \apj, 465, 79

\noindent
Soszyński, I., Udalski, A., Szymański, M.~K., et al.\ 2009, \actaa, 59, 239

\noindent
Srinivasan, S., Boyer, M.~L., Kemper, F., et al.\ 2016, \mnras, 457, 2814

\noindent
Tully, R.~B., Rizzi, L., Dolphin, A.~E., et al. 2006, AJ, 132, 729

\noindent
Valcheva, A.~T., Ivanov, V.~D., Ovcharov, E.P. et al.  2007, A\&A, 466, 501

\noindent
van de Rydt, F., Demers, S. \& Kunkel, W.~E. 1991, AJ, 102, 1301

\noindent
Weinberg, M.~D., \& Nikolaev, S. 2001, \apj, 548, 712

\noindent
Whitelock, P.~A., Feast, M.~W., van Leeuwen, F. et al. 2008, \mnras, 386, 313

\noindent
Willson, L.~A.\ 2000, \araa, 38, 573

\noindent
Wong, K.~C., Suyu, S.~H., Chen, G.~C.-F., et al. 2019, \mnras, arXiv:1907.04869

\noindent 
Wood, P.~R. 2000, PASA, 18, 18
\rm

\noindent
Wyder, T. 2003, AJ, 125, 3097

\noindent
Yuan, W., He, S., Macri, L.~M., et al.\ 2017a, \aj, 153, 170

\noindent
Yuan, W., Macri, L., He, S., et al. 2017b, ApJ, 154, 149

\noindent
Yuan, W., Macri, L., Javadi, A., et al. 2018, \apj, 156, 112 

\noindent
Zhang, B.~R., Childress, M.~J., Davis, T.~M., et al.\ 2017, \mnras, 471, 2254

\vfill\eject
\appendix

\section{Additional Galaxies}
\label{app:othergals}

We provide here discussions for  Leo~I, SagDIG, UGCA~438 and Phoenix,  which have relatively small samples of measured JAGB stars.  In addition we discuss  NGC~6822, which has a large foreground extinction. Their summary data are also tabulated in Table 1 (above) for completeness. 

\subsection{Leo I}

\noindent {\it JAGB stars: } Held et al. (2010) used WFCAM on UKIRT in Hawaii to obtain $JHK$ photometry of stars in the dwarf 
spheroidal galaxy Leo~I. Figure \ref{fig:CMDmontage} shows the resulting CMD from which we have determined an 
apparent JAGB distance modulus of 22.10 $\pm$ 0.07~(stat)~mag, using only 12 JAGB stars. Applying a Galactic foreground extinction of 
$A_J = $ 0.026~mag (NED) gives a true distance modulus of $\mu_o (JAGB) = $ 22.07 $\pm$~0.07~(stat)~mag.

\noindent {\it TRGB stars: } There are four independent TRGB distance
determination published for Leo~I. Set to a common zero point and
corrected for a foreground extinction of $A_I = $ 0.055~mag (NED)
the true moduli are: $\mu_o = $  22.05~mag (Caputo et al. 1999),
22.02~mag (Bellazzini et al. 2004), 22.19~mag (Mendez et al. 2002)
and 22.14~mag (Pietrzynski et al. 2010); and they average to
$<\mu_o (TRGB)> ~= $ 22.04 $\pm$ 0.04~(stat)~mag.

\subsection{Sagittarius Dwarf Irregular Galaxy (SagDIG)}

\noindent {\it JAGB stars: } Near-infrared $JK_s$ observations of the Sagittarius Dwarf Irregular Galaxy (Sag DIG) 
were made by Gullieuszik et al. (2007b) using the $SOFI$ instrument on the NTT at La Silla; the photometry 
for only 30 color- and magnitude-selected red $AGB$ stars were ultimately published. The CMD based on these 
data can be seen in Figure \ref{fig:CMDmontage}, from which we determine a JAGB apparent distance modulus of 24.95 $\pm$ 0.06~(stat)~mag, derived from only 19 JAGB stars.
Applying a foreground extinction correction of $A_J = $ 0.088~mag (NED) gives a true distance modulus 
of $\mu_o (JAGB) = $ 24.86 $\pm$ 0.06~(stat)~mag.

\noindent {\it TRGB stars: } Four published TRGB distances, scaled to $M_I = $ -4.05~mag and uniformly corrected for a foreground 
extinction of $A_I = $ 0.186~mag (NED) are $\mu_o = $ 25.13~mag (Karachentsev et al. 2002), 
25.24~mag (Karachentsev et al. 1999), 25.16~mag (Momany et al. 2002) and 24.96~mag (Beccari et al. 2014).
They average to $<\mu_o (TRGB)>~=~$~25.12 $\pm$ 0.14~mag.

\subsection{UGCA~438 = ESO 2323-326}

\noindent {\it JAGB stars: } Gullieuszik et al. (2008) obtained near-IR $JK$ images of UGCA~438 (UKS~2323-326) using an adaptive-optics 
system (MAD) deployed on the $VLT$. The CMD based on these 
data is shown in Figure \ref{fig:CMDmontage}, from which we determine a JAGB apparent distance modulus of 26.85 $\pm$ 0.06~(stat)~mag, based on 24 JAGB stars.

Applying a foreground extinction correction of $A_J = $ 0.010~mag (NED) gives a true distance modulus 
of $\mu_o (JAGB) = $ 24.84 $\pm$ 0.06~(stat)~mag.

\noindent {\it TRGB stars: } Six TRGB distance are available for homogenization, adopting a foreground extinction 
of $A_I = $ 0.022~mag 
and using $M_I = $ -4.05~mag for the TRGB zero point. The individual moduli are: $\mu_o = $ 
26.68~mag (Lee \& Byun 1999), 26.69~mag (McCall et al. 2012), 26.71~mag 
(Dalcanton et al. 2009), 
26.72~mag (Jacobs, et al. 2009), 26.74~mag (Tully et al. 2006) \& 26.73~mag 
(Karachentsev et al.2002). 
The averaged true TRGB distance modulus is $<\mu_o(TRGB)> ~=~ $ 26.71 $\pm$ 0.01~(stat)~mag.

\subsection{Phoenix}
\noindent {\it JAGB stars: } Menzies et al.(2008) published near-infrared ($JHK$) data for 95 of the brightest resolved stars in the Phoenix dwarf 
galaxy. In that sample there are only two JAGB stars. They have apparent $J$-band magnitudes of 16.82 and 17.11~mag
averaging to 16.96~mag. These give an apparent distance modulus of 23.16~mag. Correcting for a foreground extinction 
of $A_J$ = 0.014~mag (NED) this gives a true distance modulus of $\mu_o (JAGB) = $~23.15~mag $\pm$0.18~(stat)~mag. 

\noindent {\it TRGB stars: } Five independently determined TRGB distance moduli in NED standardized to $M_I = $ -4.05~mag and corrected 
for a foreground extinction of $A_I = $ 0.024~mag (NED) are: $\mu_o = $ 23.03~mag (Martinez-Delgado, Gallart 
\& Aparicio 1999), 23.12~mag (Held, Savianne \& Momany 1999), 23.11~mag (Battaglia et al. 2012), 23.17~mag (Hidalgo et al. 2009) and 23.28~mag (van de Rydt, Demers \& Kunkel 1991). 
Their averaged true TRGB distance modulus is $<\mu_o (TRGB)>~=~ $ 23.14 $\pm$ 0.04~(stat)~mag. 

\subsection{NGC~6822}

Finally, we consider NGC 6822, a dwarf irregular galaxy and a member of the Local Group, found at low 
Galactic latitude, and consequently subject to considerable line-of-sight Galactic extinction.

\noindent {\it JAGB stars: } Here we have the opportunity to apply the JAGB method to three independent datasets:

\noindent (1) Rich et al. (2014) recently published a multi-wavelength study of Cepheids in 
NGC~6822, in which they determined distances and reddenings. 
In it they derived an {\it true} distance modulus of 23.38 $\pm$0.05~mag,
based on the Leavitt Law applied to seven optical, near- and 
mid-infrared wavelengths. For this disk population of Cepheids, Rich 
et al. derive a value of $E(B-V) = $ 0.35 $\pm$0.04~mag for 
the total line-of-sight (Milky Way plus NGC~6822 internal) reddening
(which corresponds  to $A_J = 0.29\pm$0.03~mag.) The Galactic foreground
component in the $J$-band is found  
to be $A_J =$ 0.17~mag, using the Schlafly \& Finkbeiner (2011)
calibration given in NED. The difference between the (lower) 
foreground component and the (higher) Cepheid 
determination can plausibly be ascribed to internal extinction within 
the host galaxy, NGC~6822 itself, modulo 
the known patchiness across the line of sight and reasonable differences between the internal extinction for the very young Cepheids and for the intermediate-aged $AGB$ population, which 
could be smaller (as is seen).

Figure \ref{fig:CMDmontage} shows the observed $J$ vs $(J-H)$ CMD and 
the $J$-band luminosity function for the reddest stars 
in NGC~6822 from the Rich et al. (2014) study (side-by-side with the
other two studies discussed below). 249 stars contribute to the JAGB
luminosity function, having a mean and sigma of 18.46 and $\pm$
0.01~(stat)~mag, respectively. These give an apparent
$J$-band distance modulus and sigma on the mean of 23.66 $\pm$
0.01~(stat)~mag. 

\noindent (2) Sibbons et al. (2012) had earlier published near-infrared ($JHK$) data in a photometric study of 
the AGB population in NGC~6822, with the main focus being the subdivision of carbon-rich and oxygen-rich 
(M-Type) AGB stars. They published only reddening-corrected data where each star was dereddened 
using unique/individual line-of-sight corrections from  Schlafly \& Finkbeiner (2011). However, those 
reddening values were not included in the publication so a direct comparison with Rich et al. (2014) 
is not possible, but we can still determine a $J$-band true distance modulus from those data, with
those (implicit) reddening corrections.

Figure \ref{fig:CMDmontage} shows the observed $J$ vs $(J-K)$ CMD and the $J$-band luminosity function for the reddening-corrected 
Sibbons et al. (2012) stars in NGC~6822. In this case, 202 stars contribute to the JAGB luminosity function, 
giving a mean and sigma on the mean of 17.37 and $\pm$ 0.015~(stat)~mag, respectively. These result in a true $J$-band distance 
modulus and sigma-on-the-mean of 23.57 $\pm$ 0.02~(stat)~mag.

\noindent (3) Cioni \& Habing (2005) published near-infrared ($IJK$) data in a photometric study of 
the stellar populations in NGC~6822. 
Figure \ref{fig:CMDmontage} shows the observed $J$ vs $(J-K)$ CMD and the $J$-band luminosity function for 
25,800 stars in NGC~6822. In this case 1,443 stars contribute to the JAGB luminosity 
function, giving a mean and sigma of 17.40 and $\pm$ 0.005~(stat)~mag, respectively. These result 
in an apparent $J$-band distance modulus and sigma-on-the-mean of 23.60 $\pm$ 0.01~mag, respectively. 

Taking the mean of the three apparent moduli gives 23.61 $\pm$ 0.02~mag. Correcting for a foreground extinction 
of $A_J = $ 0.167~mag (NED) gives a true JAGB distance modulus of $\mu_o(JAGB) = $~23.44. $\pm$ 0.02~mag. 

\noindent {\it TRGB stars: } Five independently-derived TRGB distances have been published to date. 
Corrected for a foreground Galactic extinction 
of $A_I = $ 0.355~mag (NED), and using $M_I = $ -4.05~mag, the individual values are:
$\mu_o =  $ 23.50~mag (McCall et al. 2012), 23.45~mag (Cioni \& Habing 2005), 23.59~mag 
(Wyder 2003), and 23.49 (Fusco et al. 2012). 
This gives an average true modulus of $<\mu_o (TRGB)> ~=~ $ 23.52 $\pm$ 0.03~mag.

All of the differences between the Cepheid, TRGB and JAGB distance moduli can 
plausibly be accounted for by increasing the line-of-sight extinction (in addition 
to the Galactic foreground contribution) to the TRGB and JAGB stars by $A_J$ 
($A_I$) = 0.06 (0.13)~mag, internal to the host galaxy, NGC~6822.  

\end{document}